\documentclass[twocolumn,preprintnumbers,amsmath,amssymb]{revtex4}

\usepackage{graphicx}
\usepackage{dcolumn}
\usepackage{bm}

\usepackage{color}
\usepackage{placeins}

\begin{document}

\title{Optical harmonic generation on the exciton-polariton in ZnSe}


\author{Johannes Mund$^1$, Dmitri R. Yakovlev$^{1,2}$, M. A. Semina$^{2}$, and Manfred Bayer$^{1,2}$}
\affiliation{$^{1}$Experimentelle Physik 2, Technische Universit\"at Dortmund, D-44221 Dortmund, Germany}
\affiliation{$^2$Ioffe Institute, Russian Academy of Sciences, 194021 St. Petersburg, Russia}


\begin{abstract}
We study optical harmonic generation on the 1S exciton-polariton in the semiconductor ZnSe. Intense and spectrally narrow exciton resonances are found in optical second (SHG), third (THG), and fourth (FHG) harmonic generation spectra. The resonances are shifted to higher energy by 3.2~meV from the exciton energy in the linear reflectivity spectrum. Additional resonances are observed in the THG and FHG spectra and assigned to combinations of incident and backscattered photons in the crystal. Rotational anisotropy diagrams are measured and further information on the origin of the optical harmonic generation and the involved exciton states is obtained by a symmetry analysis using group theory in combination with a microscopic consideration.
\end{abstract}

\maketitle

\section{Introduction}
\label{sec_intro}

Since the discovery of second harmonic generation (SHG) in 1961 in a quartz crystal by Franken $et~al.$ \cite{Franken61}, the physics of the optical harmonic generation has developed into a mature field of both basic and applied research~\cite{Shen, Boyd}. As a coherent process, SHG requires electronic states which can be excited by two-photon excitation and release one-photon emission. Additionally to fulfillment of the energy and wave vector conservation laws, the respective optical transitions need to be allowed in electric-dipole (ED) approximation, or in higher order, like in electric-quadrupole (EQ) and/or magnetic-dipole (MD) approximation. This makes SHG a valuable and informative tool for exciton spectroscopy in semiconductors, which delivers information not available from linear optical spectroscopy approaches~\cite{Yakovlev18}.

SHG exciton spectroscopy has been used to study various semiconductors, like GaAs~\cite{Pavlov05, Saenger06}, CdTe~\cite{Saenger06}, ZnO~\cite{Lafrentz13}, and Cu$_2$O~\cite{Mund18}. Due to the resonant enhancement by the exciton states, the SHG signals show up as narrow lines which, at low temperatures, often are stronger than the crystallographic signals which appear also for off-resonance conditions. In these studies, magnetic and electric fields were used to reduce the symmetries of the exciton states, which gives rise to various field-induced SHG mechanisms~\cite{Lafrentz13,Farenbruch20,Brunne15}. Third harmonic generation (THG) has been also used for studying excitons, but much rarer as compared to SHG~\cite{Brunne15,Warkentin18}. 

In bulk crystals, the strong light-matter interaction leads to the formation of exciton-polaritons with characteristic dispersions given by several polariton branches~\cite{Hopfield63,Klingshirn}. Nonlinear optical spectroscopy based on multi-photon excitation can be used to measure the dispersion relations of exciton-polaritons. The pioneering study on CuCl by Fr\"ohlich $et~al.$ \cite{Froehlich71} revealed the upper polariton branch, measured directly by two-photon absorption (TPA). Shortly afterwards, similar measurements were performed by SHG in CuCl~\cite{Haueisen71,Haueisen71PRL,Haueisen73,Kramer74}, ZnO~\cite{Haueisen73,Levine75}, and CdS~\cite{Levine75}. Techniques using multi-photon excitaton, like TPA, two-photon excitation of photoluminescence, three-photon sum and difference frequency generation have typically more often been used for exciton-polariton studies compared to SHG, for a review see Ref.~\cite{Froehlich94}.

The semiconductor ZnSe has a large exciton binding energy of 20~meV leading to pronounced exciton-polariton properties \cite{Feierabend78, Mayer94}, which makes it an attractive model system for nonlinear optical spectroscopy. Its exciton-polariton dispersion is well studied by various experimental techniques: resonant Brillouin scattering \cite{Sermage79,Sermage81,Mayer94}, two-photon resonant Raman scattering \cite{Nozue81}, and two-photon excitation of photoluminescence \cite{Bogani92,Froehlich95}. We showed recently that the exciton-polariton in ZnSe can be also addressed by SHG~\cite{Yakovlev18} and THG~\cite{Warkentin18}. Note, that TPA was also used in ZnSe to study the fine structure of the 2P exciton~\cite{Sondergeld75,Minami91} and its modification in magnetic field~\cite{Sondergeld77a,Sondergeld77b,Hoelscher85} and under pressure~\cite{Reimann99}. Also, the second-order nonlinear susceptibility in ZnSe was investigated by SHG \cite{Wagner98}. 

In this paper, we report a detailed study of optical harmonic generation (SHG, THG, and forth harmonic generation (FHG) on the exciton-polariton in ZnSe. We use the recently developed technique based on spectrally broad femtosecond laser pulses and signal analysis with a high-resolution spectrometer~\cite{Mund18}. Narrow resonances are observed in the optical harmonic generation spectra that are shifted by about 3.2~meV to higher energy from the exciton resonance in the reflectivity spectrum. Rotational anisotropies of the optical harmonic generation signals are measured by rotating the linear polarizations of ingoing and outgoing photons. Their symmetries are modeled and explained in the frame of a group theory analysis and a microscopic consideration.      

The paper is organized as follows. In Section~\ref{sec_phenomenology}, a phenomenological consideration of second, third, and fourth optical harmonic generation is given. In Section~\ref{sec_experiment}, details of the experimental setup are given. The experimental results are presented in Sec.~\ref{sec_exp_results}, and the rotational anisotropies are analyzed and discussed in Sec.~\ref{sec_theory} by group theory, where also the resonances in the spectra are assigned to polariton states.

\section{Phenomenological description of optical harmonic generation}
\label{sec_phenomenology}

In semiconductors with a noncentrosymmetric crystal lattice, such as cubic ZnSe (crystallographic point group $\overline{4}3m$), the SHG process is allowed in ED approximation. The nonlinear polarization at twice the fundamental frequency, 2$\omega$, of the exciting light for the crystallographic contribution to SHG, $\mathbf{P}^{2\omega}$, reads
\begin{equation}
P_{i}^{2\omega}= \epsilon_0 \chi_{ijl}E_j^\omega E_l^\omega  ,
\label{eq:P1}
\end{equation}
where $i, j, l$ are the Cartesian indices, $\epsilon_0$ is the vacuum permittivity, $\chi_{ijl}$ is the nonlinear optical susceptibility, $E^{\omega}_{j(l)}$ are the components of the electric field $\mathbf{E}^{\omega}$ of the laser beam at the fundamental frequency $\omega$.
Equation~(\ref{eq:P1}) takes into account only the resonant and nonresonant ED contributions of the electronic states in the semiconductor at the frequencies $\omega$ and 2$\omega$. A more general approach takes into account the specifics of the material hosting exciton-polaritons. This becomes important when the SHG at frequency 2$\omega$ is in resonance with the energy of the exciton state ${\cal E}_{\mathrm{exc}}$. To account for these contributions, the effective nonlinear polarization at double frequency 2$\omega$, appearing under the excitation by the electric field of the electromagnetic wave $\mathbf{E}^\omega(\mathbf{r},t) = \mathbf{E}^\omega
\exp[{\mathrm{i}(\mathbf{k}^{\omega}\mathbf{r} -\omega t)}]$ can be written in the form
\begin{eqnarray}
P_{\mathrm{eff},i}^{2\omega}({\cal E}_{\mathrm{exc}})=\epsilon_0
\chi_{ijl}({\cal
E}_{\mathrm{exc}},\mathbf{k}_{\mathrm{exc}}) E_j^\omega E_l^\omega,
\label{eq:P2}
\end{eqnarray}
where the nonlinear susceptibility $\chi_{ijl}({\cal E}_{\mathrm{exc}}, \mathbf{k}_{\mathrm{exc}})$ takes into account the effects of spatial dispersion in the MD and EQ approximations. $\mathbf{k}_{\mathrm{exc}}=2n\mathbf{k}^{\omega}$ is the exciton wave vector, $n$ is the refractive index of light at the fundamental frequency $\omega$, and $\mathbf{k}^{\omega}$ is the wave vector of the incoming light. The nonlinear polarization from Eqs.~(\ref{eq:P1}) and (\ref{eq:P2}) leads to the SHG signal with intensity $I^{2\omega}\propto |\mathbf{P}^{2\omega}|^2$.

In the case of a resonant contribution, which includes optical transitions between the ground state of the unexcited crystal $|G\rangle$ and the exciton state $|\mathrm{Exc}\rangle$, the SHG process must be allowed both for two-photon excitation and for one-photon emission. The fulfillment of this condition depends on the symmetry of the crystal and on the geometry of the experiment. The involvement of excitons makes this picture even more complicated and interesting due to the different symmetries of the envelope wave functions of the S, P and D exciton states, which complement the symmetry given by the point group of the crystal lattice. It is worthwhile to note here that also in case where the SHG is forbidden in ED approximation, accounting for MD and/or EQ transitions can make it allowed, see e.g. Refs.~\cite{Mund18,Mund19,Farenbruch20}. In addition, the application of uniaxial mechanical stress, electric or magnetic fields, can lower the symmetry and can lead to optical harmonic generation on mixed exciton states that are forbidden otherwise. 

A similar phenomenological approach can be applied to the THG and FHG processes. In the THG case, the effective nonlinear polarization can be written as
\begin{equation}
P_{\mathrm{eff},i}^{3\omega}({\cal E}_{\mathrm{exc}})=
\epsilon_0 \chi_{ijlk}({\cal
E}_{\mathrm{exc}},\mathbf{k}_{\mathrm{exc}})
E_j^\omega E_l^\omega  E_k^\omega.
\label{eq:P33}
\end{equation}
and THG signal intensity is $I^{3 \omega}\propto |\mathbf{P}^{3\omega}|^2$. Accordingly, for the FHG case 
\begin{equation}
P_{\mathrm{eff},i}^{4\omega}({\cal E}_{\mathrm{exc}})=
\epsilon_0 \chi_{ijlkm}({\cal
E}_{\mathrm{exc}},\mathbf{k}_{\mathrm{exc}})
E_j^\omega E_l^\omega  E_k^\omega E_m^\omega.
\label{eq:P34}
\end{equation}
and FHG signal intensity is $I^{4 \omega}\propto |\mathbf{P}^{4\omega}|^2$.

Generation of optical harmonics requires the fulfillment of energy and momentum conservation. The sum of energies of $N$ ingoing photons has to be equal to the energy of the single outgoing photon, and the equivalent relations holds for the light $k$-vectors. Here $N=2,3,4$ is the harmonic order corresponding to SHG, THG, and FHG, respectively.
\begin{eqnarray}
N\hbar\omega_{N}&=&\hbar\omega^{N\omega},\\
N\textbf{k}^\omega_{N}&=&\textbf{k}^{N\omega}.
\label{eq:conservation}
\end{eqnarray}

In crystals where the phase matching condition $n^{N\omega}=n^{\omega}$ is satisfied, the laser and harmonic beams have the same phase velocity, which leads to an amplification of the generated signal intensity. The anomalous dispersion in the range of the exciton-polariton states makes it possible to satisfy the phase-matching condition in crystals~\cite{Warkentin18}, while this condition is not satisfied for the non-resonant optical harmonic generation.

\section{Experiment}
\label{sec_experiment}

\begin{figure}[h]
	\begin{center}
		\includegraphics[width=0.48\textwidth]{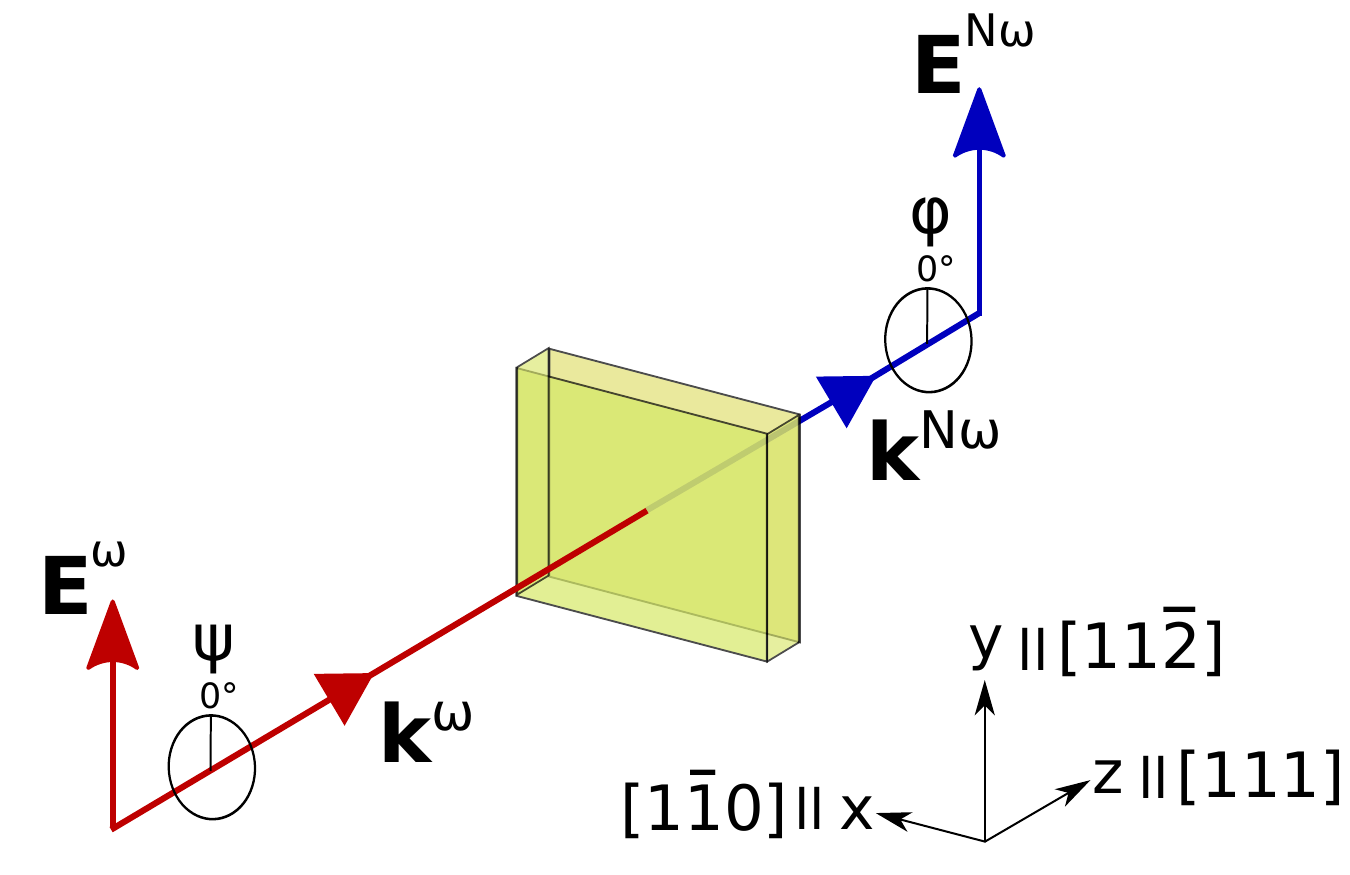}
		\caption{Sample orientation relative to the optical axis and linear polarization angles in the optical harmonic generation experiments. Ingoing light with frequency $\omega$ and electric field component $\textbf{E}^\omega$ has the wave-vector $\textbf{k}^\omega$ and polarization angle $\psi$. Generated harmonic light with frequency $N\omega$ ($N=2,3,4$) and electric field component $\textbf{E}^{N\omega}$ has the wave vector $\textbf{k}^{N\omega}$ and can be detected at polarization angle $\varphi$.}
		\label{pic.setup_and_sample}
	\end{center}
\end{figure}

We use a recently developed technique for exciton spectroscopy based on optical harmonic generation with 200~fs laser pulses and high spectral resolution. The technique is described in detail in Ref.~\cite{Mund18}. In Figure~\ref{pic.setup_and_sample}, the sample orientation relative to the optical axis as well as the linear polarization angles of the fundamental and harmonics light are specified.

The studied ZnSe bulk sample was grown by the Bridgman method. The sample was cut such that we can orient its crystal axes in the following way relative to the chosen coordinate axes:  $[111] \parallel z$ (thickness: $2586~\mu$m), $[11\bar{2}] \parallel y$ ($4475~\mu$m) and $[1\bar{1}0] \parallel x$ ($2458~\mu$m), see Fig.~\ref{pic.setup_and_sample}. For optical measurements, the sample is kept in a bath cryostat at a temperature of $T=5~\textrm{K}$ in contact with cold helium gas.

The pump laser in our setup emits pulses of 150~fs duration at a repetition rate of 30~kHz. It pumps optical parametric amplifiers (OPA) of which one emits pulses of 3.3~ps duration and a full width at half maximum (FWHM) of about 1~meV. The other OPA emits pulses of 200~fs duration and FWHM of about 10~meV. The OPA photon energy can be tuned in the range of relevance for the optical harmonic generation of about $E_g/N$, where $E_g=2.82$~eV is the band gap energy of ZnSe at cryogenic temperature. The energy per pulse is set to $0.1-1.0~\mu$J for both OPAs depending on the harmonic order to be measured.

The laser beam hits the ZnSe sample surface being parallel to the [111] crystal direction under normal incidence. It is focused into a spot with size of about 100~$\mu$m. The signals are detected by the combination of a spectrometer and a silicon charge-coupled device (CCD) camera. The 1-m Spex 1704 spectrometer has a $10\times10$~cm$^2$ sized grating with 1200-grooves/mm. The spectral resolution of the system in the energy range of the ZnSe band gap is 30~$\mu$eV. Further information on the detection system can be found in \cite{Farenbruch20}.

With a Glan-Thompson polarizer and a half-wave plate, the linear polarization of the incoming and outgoing light can be varied continuously and independently. One can thus detect the signals for any chosen polarization of $\textbf{E}^\omega$ or $\textbf{E}^{N\omega}$ and, therefore, measure the rotational anisotropy diagrams of the optical harmonics. In this paper, we measure these anisotropies for either parallel ($\textbf{E}^\omega\parallel\textbf{E}^{N\omega}$) or crossed ($\textbf{E}^\omega\perp\textbf{E}^{N\omega}$) linear polarizations of the laser and signal light.

In order to have information on the properties of the exciton-polaritons in the studied sample we use linear optical spectroscopy. For measuring the reflectivity spectrum a white-light lamp is used to illuminate the sample.

Optical harmonic generation spectra are measured by exciting the sample with laser pulses emitted by the fs-OPA. With these spectrally broad fs-pulses the whole spectral range around the 1S exciton can be excited by setting the central output photon energy of the OPA to ${\cal E}_\textrm{1S}/N$ with corresponding $N$ for SHG, THG, or FHG. Here ${\cal E}_\textrm{1S}$ is the energy of the exciton-polariton signal in the optical harmonic generation spectra. The spectral resolution achieved in the experiments with fs-pulses depends on the detection system, as described above.

A two-photon photoluminescence excitation (2P-PLE) spectrum is recorded by detecting the photoluminescence (PL) intensity at the fixed energy of $2.68956$~eV,  whereas the laser photon energy is tuned in the spectral range of the ${\cal E}_\textrm{1S}$ exciton state by scanning the photon energy of the ps-OPA. In this scanning regime, the spectral resolution depends on the OPA linewidth.

\section{Experimental Results}
\label{sec_exp_results}

\begin{figure}[h]
	\begin{center}
		\includegraphics[width=0.48\textwidth]{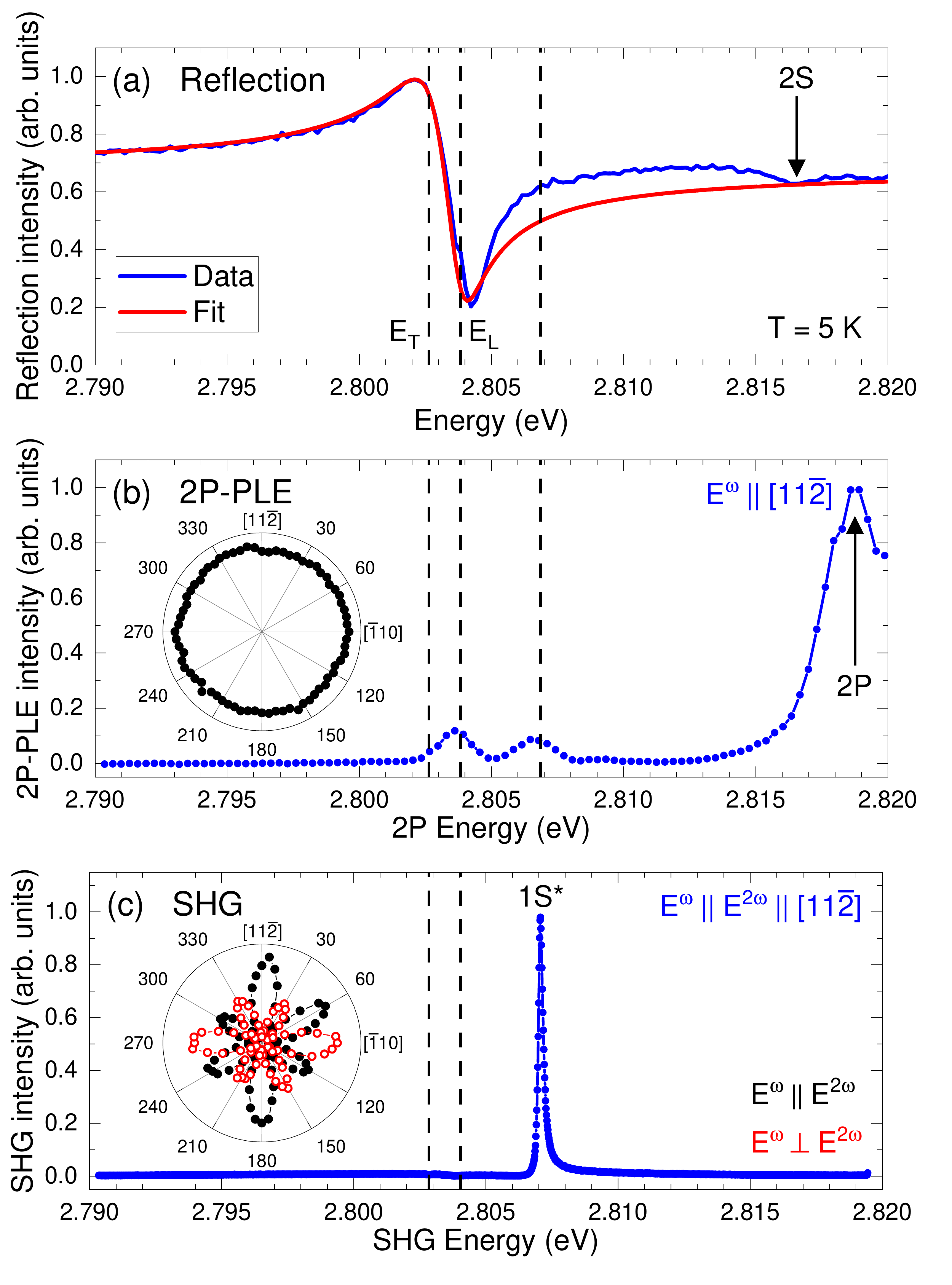}
		\caption{Comparison of: (a) white-light reflectivity spectrum, (b) 2P-PLE spectrum and (c) SHG spectrum of the 1S exciton resonance with the fundamental laser photon energy set to $\hbar\omega=1.402$~eV. In (a) the experimental spectrum (blue) is fitted (red) by the exciton-polariton approach described in the text. The inset in (b) shows the 2P-PLE intensity in dependence of the excitation polarization at energy $2\hbar\omega=2.8068$~eV. The inset in (c) shows the rotational anisotropy of the SHG intensity at resonance 1S* for parallel $\textbf{E}^\omega\parallel\textbf{E}^{2\omega}$ (filled black dots) and crossed $\textbf{E}^\omega\perp\textbf{E}^{2\omega}$ (open red dots) polarization configuration.}
		\label{pic.ZnSe_1S_WL_2PLE_SHG_111}
	\end{center}
\end{figure}

In Figure~\ref{pic.ZnSe_1S_WL_2PLE_SHG_111}, three spectra in the energy range of the 1S exciton in ZnSe are shown. The reflectivity spectrum (blue) is plotted in Fig.~\ref{pic.ZnSe_1S_WL_2PLE_SHG_111}(a). The fit (red) was made within the exciton-polariton model with taking into account the spatial dispersion in the "dead" layer approximation \cite{Pekar58, Hopfield63,Sell73}. The fit allows to obtain the values of the exciton parameters: $E_\textrm{T}=2.80268$~eV, $E_\textrm{L}=2.80389$~eV, $\hbar\omega_\textrm{LT}=1.21$~meV, and $\hbar\Gamma=1.2$~meV. These values are in accordance with literature data \cite{Feierabend78, Mayer94, Froehlich95}. We used $\varepsilon=5.5$ as the background dielectric constant and $M\approx0.8m_0$ as exciton translational mass. Furthermore, the 2S exciton state is resolved at ${\cal E}_\textrm{2S}=2.81648$~eV. The energies $E_\textrm{T}$ and $E_\textrm{L}$ are indicated in Figs.~\ref{pic.ZnSe_1S_WL_2PLE_SHG_111}(b) and \ref{pic.ZnSe_1S_WL_2PLE_SHG_111}(c) by the dashed lines.

The 2P-PLE spectrum in Fig.~\ref{pic.ZnSe_1S_WL_2PLE_SHG_111}(b) is detected at an energy of 2.68956~eV, while scanning the ps-OPA. The spectrum reveals two peaks at $E_\textrm{L}$ and at a higher energy of 2.8068~eV. Approaching the ZnSe band gap, another strong peak is seen at ${\cal E}_\textrm{2P}=2.8188$~eV, which can be assigned to the 2P exciton state allowed for two-photon excitation \cite{Sondergeld75,Hoelscher85}. In the inset of Fig.~\ref{pic.ZnSe_1S_WL_2PLE_SHG_111}(b), the signal intensity in dependence of the polarization angle of the incoming photons $\textbf{E}^\omega$ is presented for $2\hbar\omega=2.8068$~eV, which shows a completely isotropic shape.

The SHG spectrum is shown in Fig.~\ref{pic.ZnSe_1S_WL_2PLE_SHG_111}(c). The intense and narrow peak, labeled as 1S*, at ${\cal E}_\textrm{1S*}=2.80707$~eV dominates the spectrum. It is also observed in the 2P-PLE spectrum in Fig.~\ref{pic.ZnSe_1S_WL_2PLE_SHG_111}(b). In SHG, the FWHM of this line $\Gamma_\textrm{1S*}=180~\mu$eV is not limited by the resolution of our detection system, but by inherent properties of the exciton. The rotational anisotropy of the 1S* line is depicted in the inset. The full black and open red dots correspond to the parallel $\textbf{E}^\omega\parallel\textbf{E}^{2\omega}$ and crossed $\textbf{E}^\omega\perp\textbf{E}^{2\omega}$ configuration, respectively. In both configurations, a very pronounced modulation of the anisotropy can be seen, which is characteristic for the SHG signals, in contrast to the 2P-PLE. The modulation originates from the crystal symmetry and is specific for the chosen crystal orientation, details will be discussed below.

\begin{figure}[h]
	\begin{center}
		\includegraphics[width=0.48\textwidth]{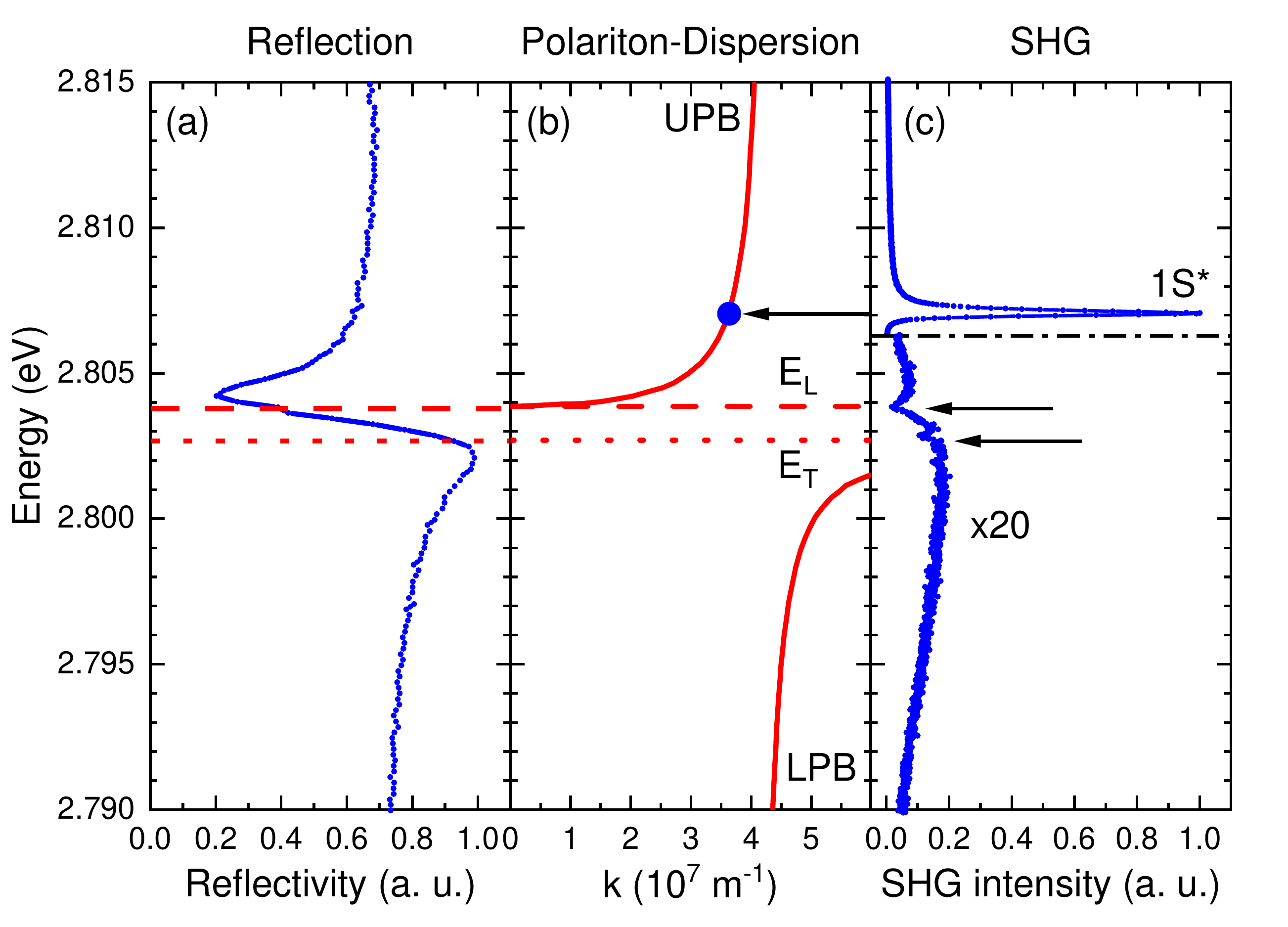}
		\caption{Assignment of the features in the (a) reflectivity and (c) SHG spectrum to the exciton-polariton dispersion (b) \cite{Froehlich95}. The solid lines in panel (b) refer to the UPB and LPB. The longitudinal exciton is marked by the dashed line and the transversal exciton by the dotted line. Note that in panel (c) the SHG intensity in the energy range below the dash-dotted line is magnified by a factor of 20. 
		}
		\label{pic.ZnSe_1S_2PLE+Disp+SHG_111}
	\end{center}
\end{figure}


We compare the reflectivity and the SHG spectrum with the exciton-polariton dispersion of the 1S exciton in ZnSe, taken from Ref.~\cite{Froehlich95}, see Fig.~\ref{pic.ZnSe_1S_2PLE+Disp+SHG_111}, to assign the observed features to particular states. In Figure~\ref{pic.ZnSe_1S_2PLE+Disp+SHG_111}(b), the solid lines show the exciton-polariton dispersion consisting of the upper polariton branch (UPB) and the lower polariton branch (LPB)~\cite{Klingshirn}. The dashed line marks the longitudinal exciton energy, whereas the dotted line indicates the transversal exciton energy. The intense 1S* resonance in the SHG spectrum is located in the UPB 3.2~meV above $E_\textrm{L}$. Furthermore, the SHG spectrum reveals two dips at $E_\textrm{L}=2.80389$~eV and $E_\textrm{T}=2.80268$~eV, indicating the longitudinal and transversal exciton, respectively. On the one hand, the longitudinal state can be excited by two photons, as was observed in the 2P-PLE spectrum, but does not emit light at the same energy leading to a dip in the SHG spectrum. On the other hand, the broad SHG signal below $E_\textrm{T}$ is related to nonresonant SHG on the LPB.

\begin{figure}[h]
	\begin{center}
		\includegraphics[width=0.48\textwidth]{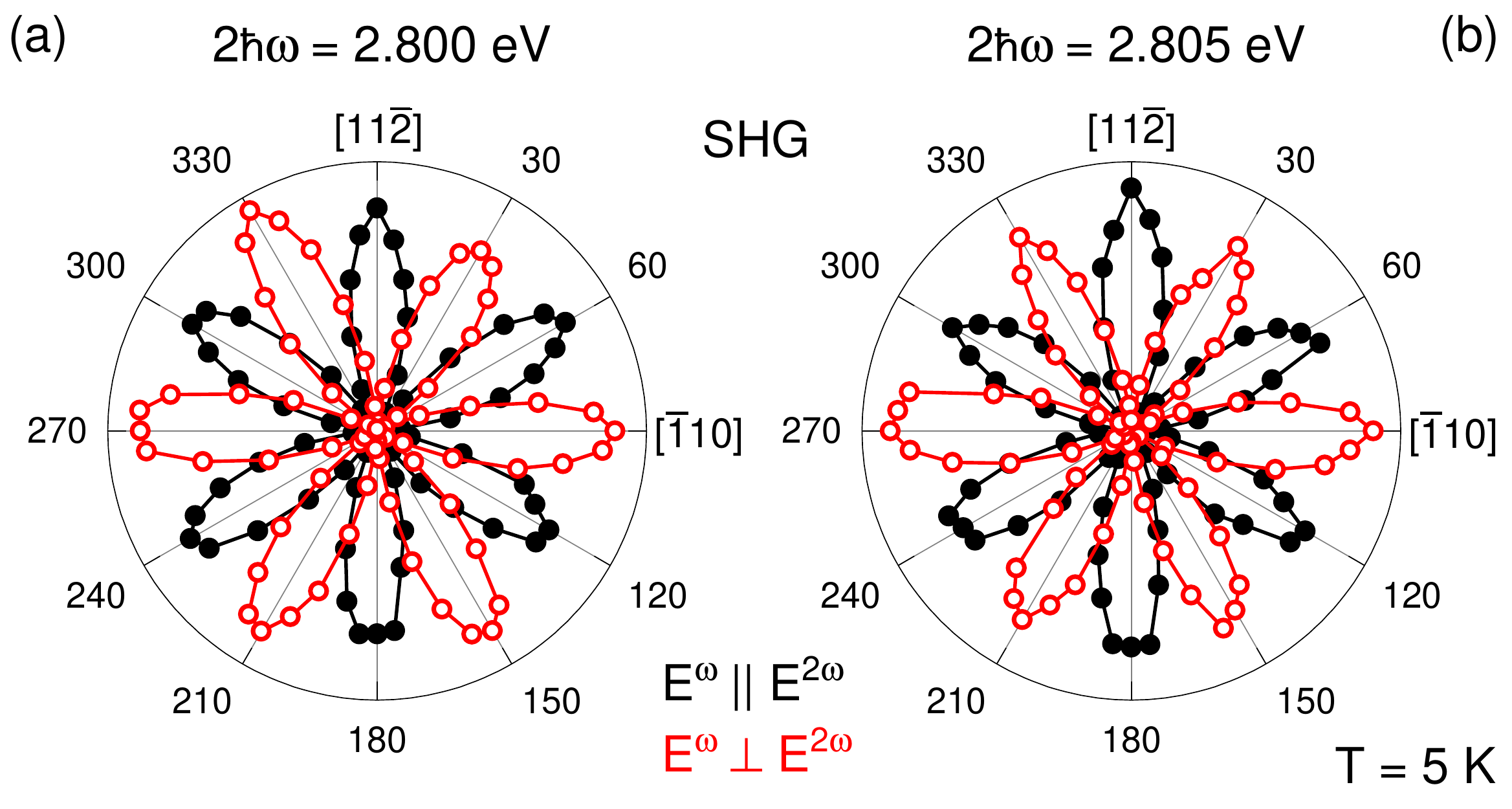}
		\caption{SHG rotational anisotropies at the energies (a) 2.800~eV and (b) 2.805~eV. Full black and open red dots give data for parallel $\textbf{E}^\omega\parallel\textbf{E}^{2\omega}$ and crossed $\textbf{E}^\omega\perp\textbf{E}^{2\omega}$ configuration, respectively.}
		\label{pic.ZnSe_1S_SHG_111_Anis}
	\end{center}
\end{figure}

In Figure~\ref{pic.ZnSe_1S_SHG_111_Anis}, we show SHG rotational anisotropies measured at the energy of 2.805~eV, which is between the 1S* energy and $E_\textrm{L}$, and at 2.800~eV corresponding to the broad SHG signal at energies below $E_\textrm{T}$. Both anisotropies are very similar to each other and show a sixfold pattern with lobes of equal intensity. 

\begin{figure}[h]
	\begin{center}
		\includegraphics[width=0.48\textwidth]{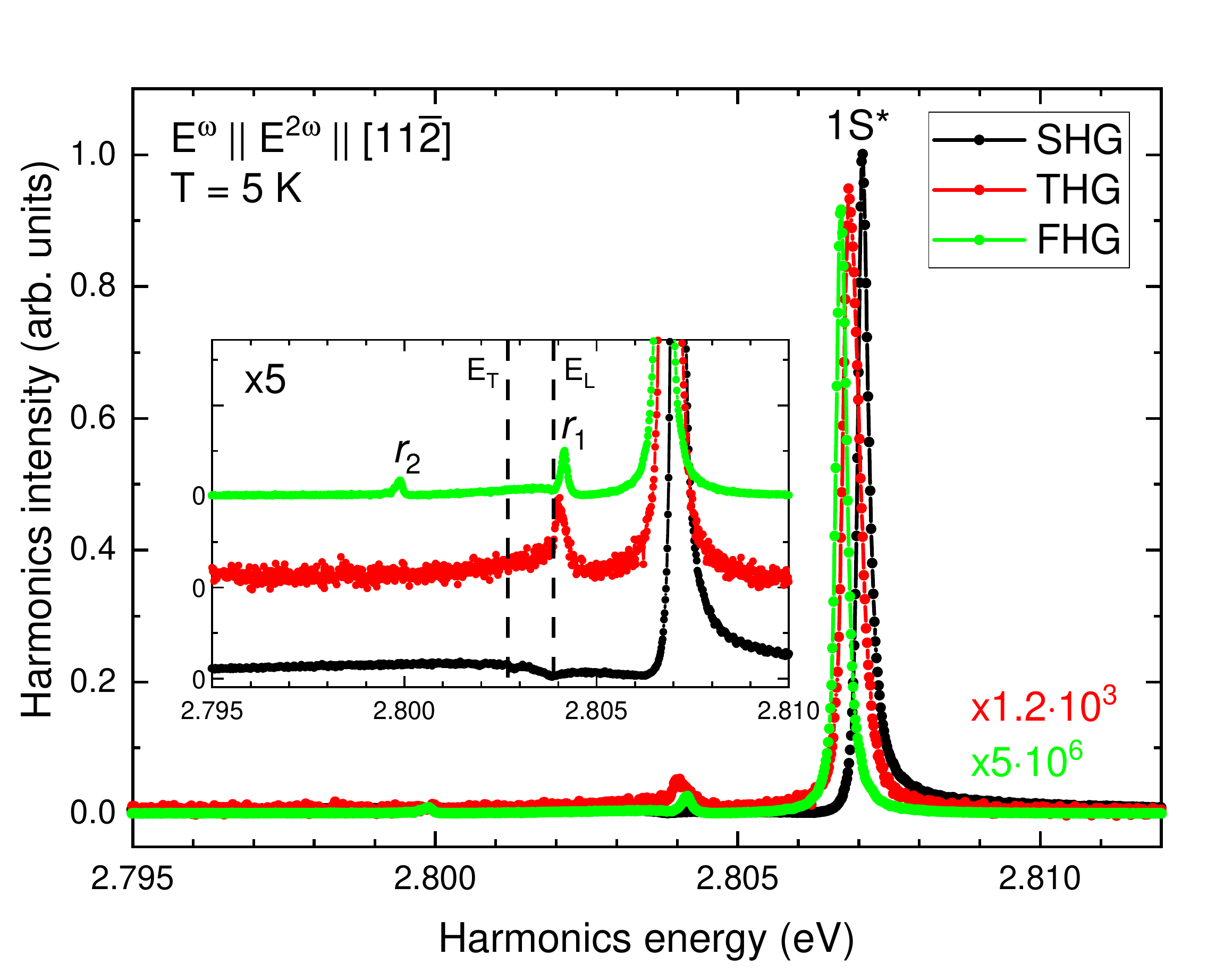}
		\caption{SHG (black), THG (red), and FHG (green) spectra of the 1S* exciton resonance. In the inset the spectra are magnified by a factor of 5 to highlight the peaks $r_1$ and $r_2$. Note the scaling factors of $1.2\cdot10^3$ for the THG and $5\cdot10^6$ for the FHG spectra to match the 1S* resonance SHG intensity.}
		\label{pic.ZnSe_1S_SHG-4HG_111}
	\end{center}
\end{figure}

In Figure~\ref{pic.ZnSe_1S_SHG-4HG_111}, the THG and FHG spectra are presented along with the SHG spectrum. With increasing harmonic order, the 1S* resonance is shifting to lower energy, see Table~\ref{tab.peak_energies}. Furthermore, as shown in the inset, the additional small peaks $r_1$ and $r_2$ appear on the low energy side of the 1S* resonance. The $r_1$ resonance is not related to the longitudinal exciton state which is symmetry forbidden for one-photon emission. Further, $r_1$ is located at slightly higher energy, which indicates its origin on the UPB.

\begin{table}[htbp]
	\begin{center}
		\caption{Summary of the peak energies of the 1S*, $r_1$, and $r_2$ lines taken from the SHG, THG and FHG spectra in Fig.~\ref{pic.ZnSe_1S_SHG-4HG_111} in comparison to the energies of the longitudinal and transversal exciton extractad from reflectivity. All values are given in units of (eV).}
		\label{tab.peak_energies}
		\begin{tabular}{|c|c|c|c|c|c|} \hline
			Measurement & 1S* & $r_1$ & $r_2$ & $E_\textrm{T}$ & $E_\textrm{L}$\\ \hline
			Reflectivity & & & & 2.80268 & 2.80389\\
			SHG & 2.80707 & & & &\\
			THG & 2.80685 & 2.80404 & & &\\
			FHG & 2.80671 & 2.80417 & 2.79992 & &\\ \hline
		\end{tabular}
	\end{center}
\end{table}

\begin{figure}[hbt]
	\begin{center}
		\includegraphics[width=0.48\textwidth]{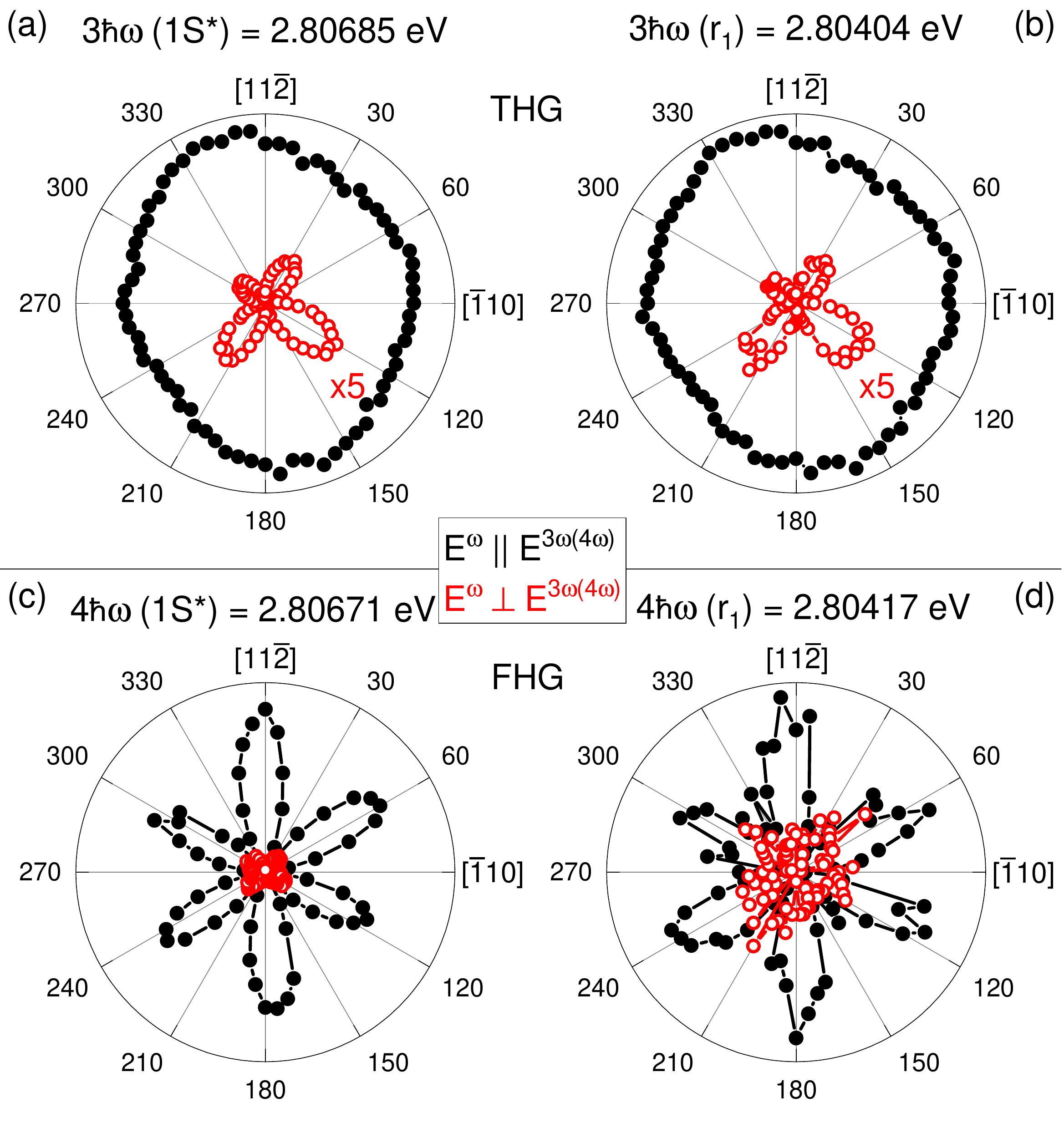}
		\caption{Rotational anisotropies of the resonances 1S* (a) and $r_1$ (b) in THG and in FHG (c) and (d). Full black and open red dots represent data for parallel $\textbf{E}^\omega\parallel\textbf{E}^{2\omega}$ and crossed $\textbf{E}^\omega\perp\textbf{E}^{2\omega}$ polarization configuration, respectively. Note that the crossed signal in the THG anisotropies is magnified by a factor of 5 for better visibility.}
		\label{pic.ZnSe_THG+4HG_1S+r1_111_Anis}
	\end{center}
\end{figure}

The rotational anisotropies of the 1S* and $r_1$ resonances for THG and FHG are shown in Fig.~\ref{pic.ZnSe_THG+4HG_1S+r1_111_Anis}. In THG, the anisotropy shape is almost isotropic with only a slight modulation for the parallel configuration. The intensity in the crossed configuration is much weaker and reveals a fourfold pattern with different lobe intensities. In FHG, the parallel anisotropy shows a sixfold pattern, resembling the one in SHG, compare with Fig.~\ref{pic.ZnSe_1S_SHG_111_Anis}. Again, the signal in crossed configuration is much weaker and does not exhibit any pronounced modulated pattern. Note, that the anisotropies of the $r_1$ resonance in Fig.~\ref{pic.ZnSe_THG+4HG_1S+r1_111_Anis} have the same shape as those of the 1S* line.

It is worthwhile to remind that THG energy shifts of about 0.8~meV and 1~meV relative to the bare exciton position were also seen in GaAs and CdTe, respectively \cite{Warkentin18}. However, ZnSe with 20~meV exciton binding energy \cite{Woerz97} is particularly suited to observe a clear polariton-shift. Similar measurements on Cu$_2$O with a 1S exciton binding energy of 150~meV \cite{Uihlein81} do not show a shift due to its different crystal symmetry and band structure. Thus, one-photon optical transitions are "forbiddden", e.g. only quadrupole-allowed, and excitons have lower oscillator strengths.

\section{Theory}
\label{sec_theory}

In this section, we introduce first the exciton-polariton concept to explain the spectral shift of the 1S* line in the optical harmonic spectra with respect to the exciton resonance in the reflectivity spectrum. Afterwards, we turn to a general symmetry analysis of the optical nonlinearities in ZnSe in terms of group theory. We simulate the expected rotational anisotropies for 2P-PLE and optical harmonic generation. These are explicitly given for light directed along the [111] crystal axis.  

\subsection{Exciton-Polaritons}

In semiconductors, strong light-matter interaction results in the formation of exciton-polaritons with a specific energy dispersion, modifying not only the linear optical properties, but also the nonlinear ones and thus the spectra of optical harmonic generation. Recently, the microscopic theory of THG on exciton-polaritons was developed for zinc-blende semiconductors and was used to describe the THG enhancement in magnetic field~\cite{Warkentin18}. 

\begin{figure}[h]
	\begin{center}
		\includegraphics[width=0.48\textwidth]{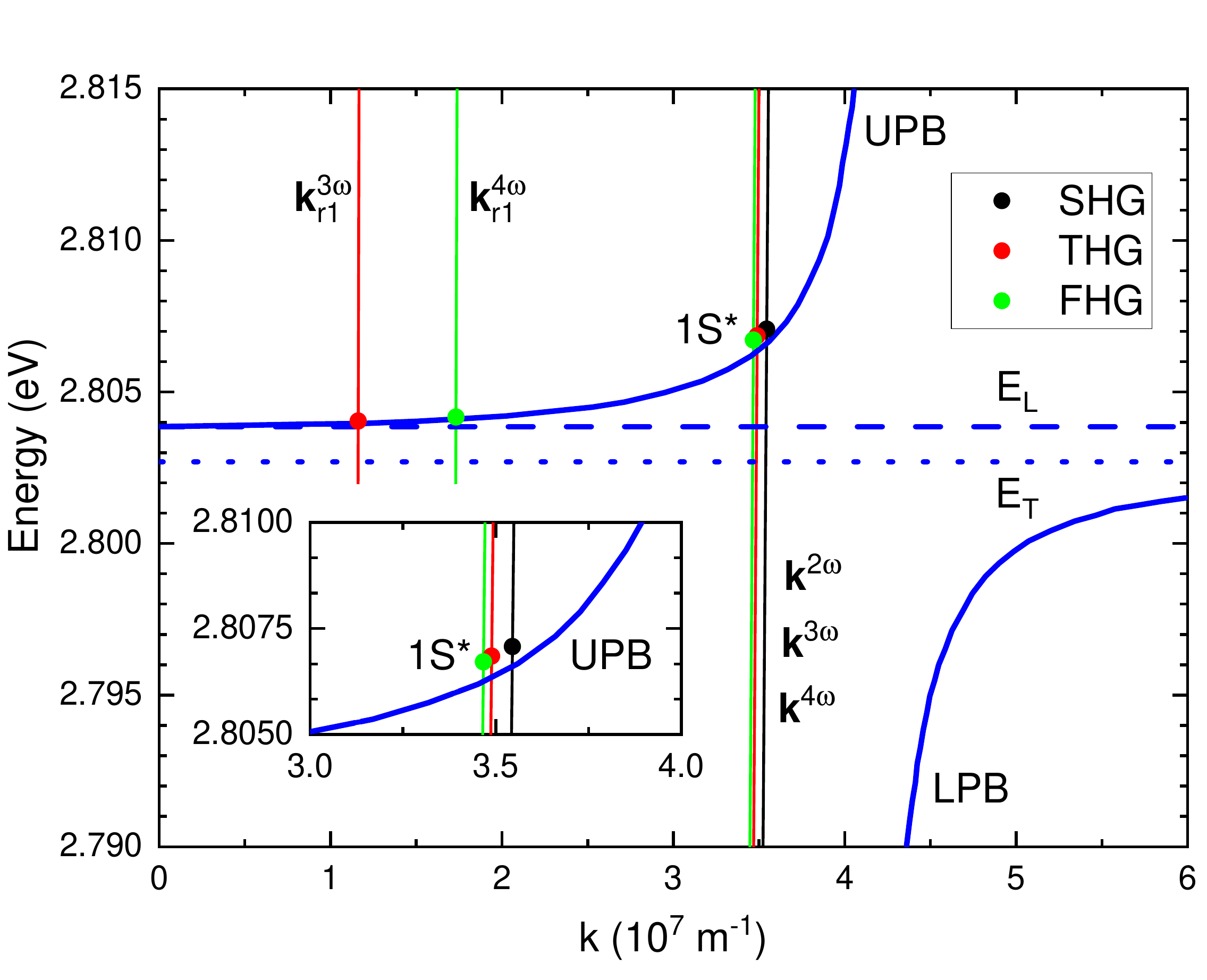}
		\caption{Comparison of the 1S exciton-polariton dispersion curves consisting of the UPB and LPB (solid blue lines) and the peak positions in the harmonic measurements, shown in Fig.~\ref{pic.ZnSe_1S_SHG-4HG_111}. Also the longitudinal and transversal exciton energies (dashed and dotted blue lines) are shown. The points give the resonance energies in SHG (black), THG (red) and FHG (green). The almost vertical lines give the dispersions of the fundamental light in the corresponding harmonic measurement. The inset shows a zoom of the dispersions around the 1S* resonances.} 
		\label{pic.ZnSe_Polariton-Dispersion}
	\end{center}
\end{figure}

In Figure~\ref{pic.ZnSe_Polariton-Dispersion}, the dispersion of the 1S exciton-polariton in ZnSe with the UPB and LPB is plotted~\cite{Froehlich95}. The longitudinal, $E_\textrm{L}$, and transversal, $E_\textrm{T}$, exciton energies are given by the dashed and dotted lines, respectively. A resonant signal in the optical harmonic generation spectra corresponds to the energy, where the linear light dispersion $N\textbf{k}^\omega_{N}$ crosses the UPB, see the colored dots in Fig.~\ref{pic.ZnSe_Polariton-Dispersion}. The light wave vectors $\textbf{k}^\omega_{N}$ of the incoming photons can be calculated knowing the index of refraction at the respective energy $n(\hbar\omega_{N})$. 
\begin{equation}\label{eq.k}
\textbf{k}^\omega_{N}=\frac{n(\hbar\omega_{N})e}{c\hbar}\hbar\omega_{N},
\end{equation}
where $e$ is the elementary charge and $c$ the speed of light. We take the $n(\hbar\omega_{N})$ values for ZnSe from Table~8 in Ref.~\cite{Li84}. Here, the data are given for $T=93$~K and are expected to be slightly larger for $T=5$~K.

The $n(\hbar\omega_{N})$ values, used by us, are given in Table~\ref{tab.k-values} together with the results for $\textbf{k}^\omega_{N}$ of Eq.~\eqref{eq.k} for the specific photon energies of the resonances in SHG, THG, and FHG. 

\begin{table}[htbp]
	\begin{center}
		\caption{Summary of fundamental photon energy $\hbar\omega_{N}$, corresponding index of refraction $n(\hbar\omega_{N})$ \cite{Li84}, peak energies $\hbar\omega^{N\omega}$ from Fig.~\ref{pic.ZnSe_1S_SHG-4HG_111} and $\textbf{k}^{N\omega}$ values for SHG, THG and FHG.}
		\label{tab.k-values}
		\begin{tabular}{|c|c|c|c|c|} \hline
			 & & & & \\
			& $\hbar\omega_{N}$~(eV) & $n(\hbar\omega_{N})$ & $\hbar\omega^{N\omega}$~(eV) & $N\textbf{k}^{\omega}_N$ ($10^7 \textrm{m}^{-1}$)\\ \hline
			SHG & 1.40354 & 2.492 & 2.80707 (1S*) & 3.545\\
			THG & 0.93562 & 2.453 & 2.80685 (1S*) & 3.489\\
				& 0.93468 & 2.453 & 2.80404 (r$_1$) & 3.485\\
			FHG & 0.70168 & 2.438 & 2.80671 (1S*) & 3.467\\
				& 0.70104 & 2.438 & 2.80417 (r$_1$) & 3.464\\
				& 0.69998 & 2.438 & 2.79992 (r$_2$) & 3.459\\ \hline
		\end{tabular}
	\end{center}
\end{table}

The harmonic resonances are given as the colored dots in Fig.~\ref{pic.ZnSe_Polariton-Dispersion}, together with the corresponding fundamental photon dispersions. The resonances $r_1$ from the THG and FHG spectra are plotted at one third and at half of the calculated $k$-values, respectively. They give access to the UPB at small $k$-values whereas the 1S* resonances are located at larger $k$ on the polariton dispersion \cite{Froehlich71}. The explanation is that the $r_1$ line arises from combinations of the ingoing photons and those which are reflected from the sample backside. Therefore, one observe the $r_1$ line in THG at
\begin{equation}
\textbf{k}^{3\omega}_\textrm{r1}=2\textbf{k}^\omega_3-\textbf{k}^\omega_3=\textbf{k}^\omega_3
\end{equation}
and in FHG at
\begin{equation}
\textbf{k}^{4\omega}_\textrm{r1}=3\textbf{k}^\omega_4-\textbf{k}^\omega_4=2\textbf{k}^\omega_4.
\end{equation}

The 1S* resonance of each harmonic spectrum is plotted onto the dispersion of the corresponding $k$-values: $\textbf{k}^{2\omega}$, $\textbf{k}^{3\omega}$, and $\textbf{k}^{4\omega}$. For better visibility a zoom of the plot around the 1S* line positions is shown in the inset of Fig.~\ref{pic.ZnSe_Polariton-Dispersion}. 

The $r_2$ line is observed only in the FHG spectrum, but does not show up in SHG and THG. Its origin is not clear and requires further investigations. One can only say that it cannot be associated with Brillouin scattering from the UPB to the LPB, as for the spectral energy of 1S* in ZnSe the expected shift is about 2.5~eV, see Fig.~5 in Ref.~\cite{Sermage81}, while in our experiment the $r_2$ line is shifted from 1S* by 6.8~meV.

\subsection{Symmetry analysis}

ZnSe crystallizes in the zinc blende structure of point group $T_d$. The structure is not centrosymmetric so that parity is not a good quantum number. SHG is allowed in ZnSe in ED approximation. The expected rotational anisotropy for nonlinear optical processes can be calculated when the symmetries of the involved electronic states are known. We take the necessary information from the tables of Koster $et~al.$ \cite{Koster}.

The symmetry of an exciton, $\Gamma_\textrm{exc}$, is determined by the tensor product of three irreducible representations:
\begin{equation}
\Gamma_\textrm{exc}=\Gamma_\textrm{CB}\otimes\Gamma_\textrm{VB}\otimes\Gamma_\textrm{env}.
\end{equation}
$\Gamma_\textrm{CB}$ and $\Gamma_\textrm{VB}$ give the irreducible representations of the electron in the conduction band and of the hole in the valence band, respectively. $\Gamma_\textrm{env}$ denotes the symmetry of the exciton envelope. In ZnSe, the lowest conduction band (s-orbitals of Zn) has $\Gamma_1$ symmetry. If the spin is included, an electron in the conduction band is represented by $\Gamma_6$ symmetry. The uppermost valence band (p-orbitals of Se) has $\Gamma_5$ symmetry which transforms to $\Gamma_8$ symmetry by spin. For the 1S exciton, the spherical envelope is of $\Gamma_1$ symmetry resulting in
\begin{equation}\label{eq.Gamma1S}
\Gamma_\textrm{1S}=\Gamma_\textrm{6}\otimes\Gamma_\textrm{8}\otimes\Gamma_\textrm{1}=\Gamma_3\oplus\Gamma_4\oplus\Gamma_5.
\end{equation}
The symmetry $\Gamma_3$ belongs to the pure triplet paraexciton, whereas $\Gamma_4$ and $\Gamma_5$ correspond to states with singlet orthoexciton admixture.

In the point group $T_d$, the photon dipole operator, $O_D$, transforms as the irreducible representation $\Gamma_5$. Thus, one photon can be absorbed by exciting a $\Gamma_5$ state, or can be emitted when a $\Gamma_5$ exciton recombines. If more than one photon is involved in the excitation process, states with other symmetries can be excited. In the case of SHG, the first photon excites virtually a $\Gamma_5$ state, whereas the second photon induces a transition from this intermediate state to a final one with $\Gamma_1$, $\Gamma_3$, or $\Gamma_5$ symmetry. Possible final states for single- and multi-photon excitation (up to four photons) are
\begin{eqnarray}
1\,\textrm{photon}&:&\,\Gamma_5\\\label{eq.2photonSym}
2\,\textrm{photons}&:&\Gamma_1\oplus\Gamma_3\oplus\Gamma_5\\\label{eq.3photonSym}
3\,\textrm{photons}&:&\Gamma_1\oplus\Gamma_3\oplus2\Gamma_4\oplus3\Gamma_5\\\label{eq.4photonSym}
4\,\textrm{photons}&:&3\Gamma_1\oplus2\Gamma_2\oplus5\Gamma_3\oplus6\Gamma_4\oplus7\Gamma_5
\end{eqnarray}

For modeling the rotational anisotropies we need to consider the photons in more detail. The ingoing photons are described by their wave-vector $\textbf{k}^\omega_N$ and polarization of $\textbf{E}^\omega_N$. The components $u_N$, $v_N$, and $w_N$ of the electric field $\textbf{E}^\omega_N$ depend on the polarization angle $\psi$ of the incoming photons, see Fig.~\ref{pic.setup_and_sample}. The emitted photons have the wave vector $\textbf{k}^{N\omega}$ with components $k^{N\omega}_x$, $k^{N\omega}_y$, $k^{N\omega}_z$, and polarization along $\textbf{E}^{N\omega}$ ($N=2,3$ and 4 for SHG, THG, and FHG, respectively). The components $m^{N\omega}$, $n^{N\omega}$, and $o^{N\omega}$ depend on the outgoing polarization angle $\varphi$,
\begin{eqnarray}
\textbf{E}^\omega_N=\begin{pmatrix}u_N(\psi)\\v_N(\psi)\\w_N(\psi)\end{pmatrix}&,&\,\,\,\textbf{k}^\omega_N=\begin{pmatrix}k_{Nx}\\k_{Ny}\\k_{Nz}\end{pmatrix}.\\
\textbf{E}^{N\omega}=\begin{pmatrix}m^{N\omega}(\varphi)\\n^{N\omega}(\varphi)\\o^{N\omega}(\varphi)\end{pmatrix}&,&\,\,\,\textbf{k}^{N\omega}=\begin{pmatrix}k^{N\omega}_x\\k^{N\omega}_y\\k^{N\omega}_z\end{pmatrix}.
\end{eqnarray}

\subsection{Microscopic analysis}

However, the symmetry analysis is not sufficient to predict the mechanisms that lead to the observed harmonic rotational anisotropies. A microscopic analysis is necessary to evaluate the transition probabilities to different intermediate states. This is particularly important for THG and FHG. In these cases, the final exciton state can, in principle, be excited by a manifold of excitation paths due to symmetry reasons.
	
We have done such a microscopic analysis, along the approach presented for THG in Ref.~\cite{Warkentin18}. It allows to account for specifics of excitons and exciton-polaritons, which goes beyond the symmetry considerations based on crystal symmetries and group theory. Details of this analysis will be published elsewhere. Here, we will comment where the conclusions of microscopic analysis coincide with the group theory considerations, and where they bring in additional information.

\subsection{Two-photon photoluminescence excitation}

Here and below we concentrate only on dominant processes. For example, we neglect quadrupole transitions if dipole transitions are allowed. Also, we neglect dipole processes which are weaker according to microscopic analysis.

\begin{figure}[h]
	\begin{center}
		\includegraphics[width=0.48\textwidth]{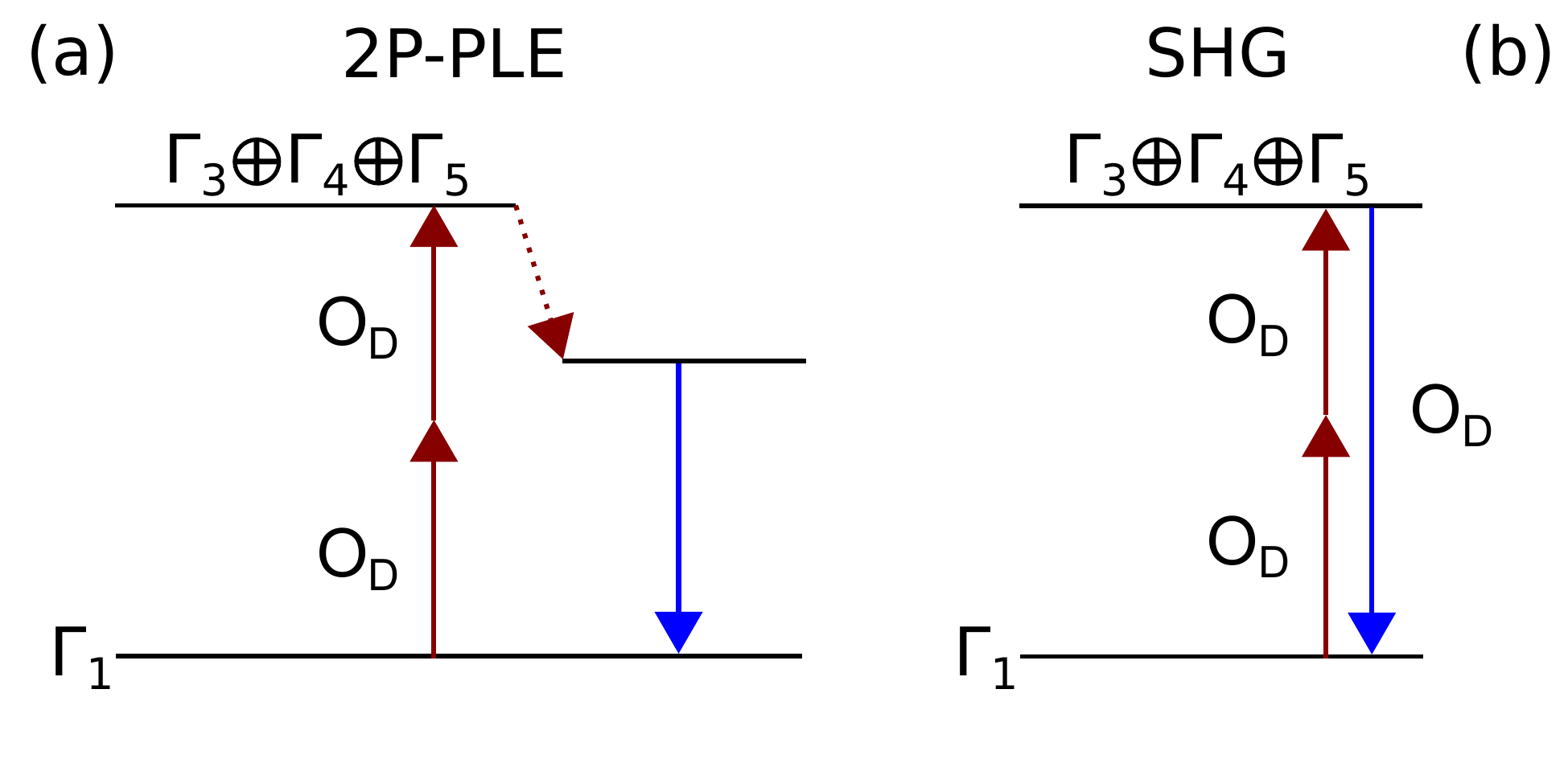}
		\caption{(a) 2P-PLE process: Excitation by two photons and emission of one photon after relaxation. (b) SHG process: Coherent excitation by two photons and emission of one photon without relaxation.}
		\label{pic.2PLE_Harmonics_Scheme}
	\end{center}
\end{figure}

The 2P-PLE process is drawn in Fig.~\ref{pic.2PLE_Harmonics_Scheme}(a). From Eqs.~(\ref{eq.Gamma1S}) and (\ref{eq.2photonSym}) we conclude that only the three components of the $\Gamma_5$ exciton can be excited by two ED transitions. States with $\Gamma_1$ and $\Gamma_3$ symmetries are not excited, because the 1S exciton does not provide a state of $\Gamma_1$ symmetry, whereas the $\Gamma_3$ paraexciton is not accessible by ED transitions. After two-photon excitation of the $\Gamma_5$ exciton it relaxes into a lower lying state and recombines with emission of one photon. Therefore, the emission in 2P-PLE is only dependent on the distinct polarization dependence of the excitation path.

The polarization-dependent two-photon ED transition is described by the operator $O_\textrm{DD}$,
\begin{equation}
O_\textrm{DD}(\textbf{E}^\omega_2, \psi)=\sqrt{2}\begin{pmatrix}v_2(\psi)w_2(\psi)\\u_2(\psi)w_2(\psi)\\u_2(\psi)v_2(\psi)\end{pmatrix}.
\end{equation}
For the present case of the light $k$-vector directed along the [111] crystal direction the polarization anisotropy of the 2P-PLE is measured in the [11$\bar{2}$]/[1$\bar{1}$0] plane. Its explicit form is 
\begin{equation}\label{eq.DD}
O_\textrm{DD}(\textbf{E}^\omega_2, \psi)=\frac{\sqrt{2}}{3}\begin{pmatrix}-\textrm{cos}(\psi)\left[\textrm{cos}(\psi)+\sqrt{3}\textrm{sin}(\psi)\right]\\\textrm{cos}(\psi)\left[-\textrm{cos}(\psi)+\sqrt{3}\textrm{sin}(\psi)\right]\\\frac{1}{2}\left[-1+2\textrm{cos}(2\psi)\right]\end{pmatrix}.
\end{equation}
The detected PL intensity is proportional to the square of $O_\textrm{DD}$
\begin{equation}\label{eq.I_2PLE}
I^\textrm{2P-PLE}\propto|O_\textrm{DD}(\psi)|^2.
\end{equation}
Note that, despite the form of Eq.~\eqref{eq.DD}, $I^\textrm{2P-PLE}$ is a constant function for all $\psi$ and thus gives an isotropic pattern for the rotational diagram.

\begin{figure}[h]
	\begin{center}
		\includegraphics[width=0.48\textwidth]{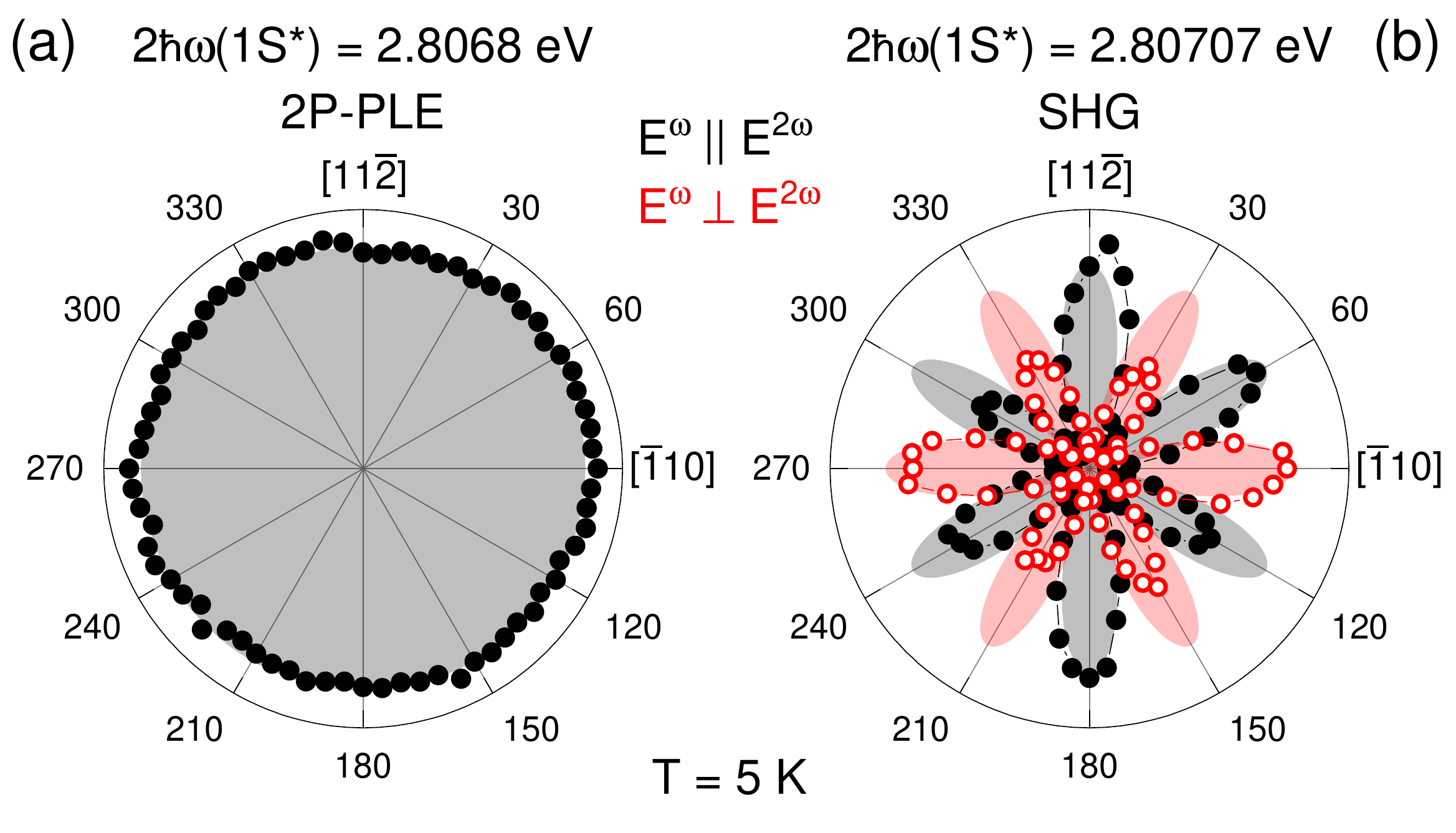}
		\caption{(a) Measured 2P-PLE data (full black dots) at the 1S* line and simulation by Eq.~(\ref{eq.I_2PLE}) (gray shaded area). (b) Full black and open red dots are measured SHG data at the 1S* resonance for parallel $\textbf{E}^\omega_2\parallel\textbf{E}^{2\omega}$ and crossed $\textbf{E}^\omega_2\perp\textbf{E}^{2\omega}$ configuration, respectively. Gray and red shaded areas represent the simulations by Eq.~(\ref{eq.I_SHG}).}
		\label{pic.2PLE+SHG_1S_111_Anis+Simu}
	\end{center}
\end{figure}

The measured rotational anisotropy of 2P-PLE on the 1S* line, shown in the inset of Fig.~\ref{pic.ZnSe_1S_WL_2PLE_SHG_111}(b), is compared with model calculations in  Fig.~\ref{pic.2PLE+SHG_1S_111_Anis+Simu}(a), where the simulation by Eq.~\eqref{eq.I_2PLE} is given by the gray shaded area.

\subsection{Second harmonic generation}
\label{subsec_theory_SHG}

For SHG, the two-photon excitation process is the same as in 2P-PLE and is followed by the one-photon emission from the same state.
Therefore, additional selection rules are due compared to 2P-PLE. The operator for one-photon emission along the [111] direction is given by
\begin{equation}
O_\textrm{D}(\textbf{E}^{2\omega}, \varphi)=\frac{1}{\sqrt{6}}\begin{pmatrix}\cos(\varphi)-\sqrt{3}\sin(\varphi)\\\textrm{cos}(\varphi)+\sqrt{3}\sin(\varphi)\\2\cos(\varphi)\end{pmatrix}.
\end{equation}
Note that the same operator is valid for the one-photon emission in THG and FHG processes considered below. The SHG intensity is calculated by
\begin{equation}\label{eq.I_SHG}
I^{2\omega}\propto|O_\textrm{DD}(\psi)O_\textrm{D}(\varphi)|^2.
\end{equation}
Thus, we expect a sixfold pattern for both parallel and crossed configuration, which are, however, rotated by $30^\circ$ with respect to each other. In Figure~\ref{pic.2PLE+SHG_1S_111_Anis+Simu}(b), the experimental SHG rotational anisotropies at the 1S* resonance are compared with the simulations according to Eq.~\eqref{eq.I_SHG}. The parallel and crossed configurations are realized by fixing $\varphi=\psi$ and $\varphi=\psi+90^\circ$, respectively.

As can be seen in Fig.~\ref{pic.2PLE+SHG_1S_111_Anis+Simu}(b), the SHG signal, measured on the 1S* line, shows general agreement, but also has some deviations from the modeling. In particular, the sixfold pattern of the simulation is reproduced, but with varying intensities of the individual lobes. On the one hand, this finding can be explained by the potential presence of residual strain in the sample. Strain can provide a splitting of the ideally threefold degenerate $\Gamma_5$ exciton. Then interference of the emission from the two transversal and the longitudinal exciton state has to be taken into account, as shown for Cu$_2$O \cite{Mund19}. The parameters of the exciton splittings under uniaxial stress in ZnSe can be found in Ref.~\cite{Froehlich95}. On the other hand, also a small deviation of $\textbf{k}^\omega_N$ from the [111] direction can induce mixing of the exciton states. Note that the 2P-PLE rotational anisotropy is not affected by interference because prior to emission the excitation relaxes into an energetically lower state. During this process the information about $k$-vector and polarization of the excitation is not conserved and, therefore, the coherence is lost.

A second possible reason for the mismatch between the SHG experiment and modeling can be photon processes beyond the ED approximation. Additionally to the considered ED transitions, EQ and/or MD transitions may contribute to SHG \cite{Mund18, Farenbruch20}. For example, the $\Gamma_5$ exciton can be excited by two photons in ED order and emit a photon in ED or EQ order, which can interfere resulting in a modified rotational anisotropy.

The microscopic analysis coincidences with the group theory approach for SHG. There is only one path due to symmetry by which the $\Gamma_5$ 1S exciton can be excited in ED approximation. Also, there is only one independent tensor component, $\chi_\textrm{xyz}$ \cite{Klingshirn}. An in depth microscopic analysis reveals that the first photon of the two-photon excitation path in SHG virtually excites a remote band before the second photon excites the 1S exciton.

\subsection{Third harmonic generation}

\begin{figure}[h]
	\begin{center}
		\includegraphics[width=0.48\textwidth]{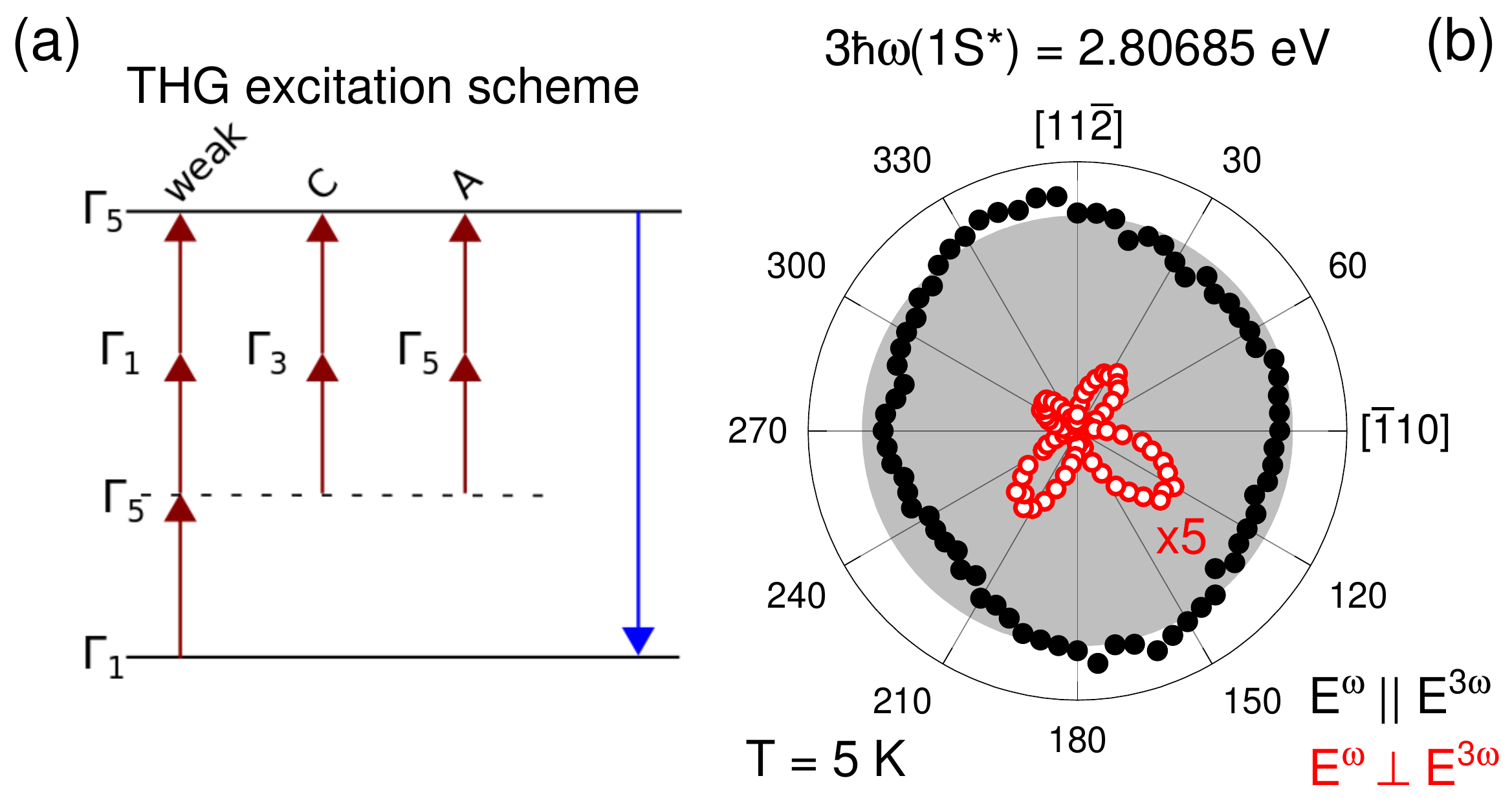}
		\caption{(a) Excitation paths for $\Gamma_5$ excitons in THG. The dashed line is a guide to the eye for the virtually excited intermediate state. Arrows represent photon transitions of $\Gamma_5$ symmetry. (b) Full black and open red dots are measured THG data at the 1S* line for parallel $\textbf{E}^\omega_3\parallel\textbf{E}^{3\omega}$ and crossed $\textbf{E}^\omega_3\perp\textbf{E}^{3\omega}$ configuration, respectively. The gray shaded area represents the simulation by Eq.~\eqref{eq.I_THG} with the ratio $\textrm{C}/\textrm{A}=1$.}
		\label{pic.THG_Scheme+Ani}
	\end{center}
\end{figure}

THG differs from SHG by the increased number of photons in the excitation process. Therefore, as shown in Eq.~\eqref{eq.3photonSym}, also states of symmetry $\Gamma_4$ can be excited by three photons. Furthermore, additional excitation paths become available for states of a certain symmetry due to the increased number of intermediate states, which are virtually excited by the photons. This number of paths is given by the coefficients in Eq.~\eqref{eq.3photonSym}. Let us discuss the case of 3$\Gamma_5$ as an example, which is illustrated in Fig.~\ref{pic.THG_Scheme+Ani}(a). By two photons, states of $\Gamma_1$, $\Gamma_3$, and $\Gamma_5$ symmetry can be virtually excited. From each of these states an additional photon of $\Gamma_5$ symmetry can excite the final 1S $\Gamma_5$ state. Thus, the increased number of excitation paths can lead to signals of different polarizations, which can interfere with each other. However, a microscopic analysis of the different paths accounting only for interband transitions and excluding transitions to remote bands gives further information about the strengths of each excitation \cite{Warkentin18}. Analysis shows that the paths via states $\Gamma_3$ and $\Gamma_5$ are strongest.

The three photon excitation operator for the light $k$-vector directed along the crystal [111]-direction is given by:
\begin{widetext}
	\begin{equation}
	O_\textrm{DDD}(\textbf{E}^\omega_3, \psi,\textrm{A},\textrm{C})=\frac{1}{6\sqrt{6}}\begin{pmatrix}2(\textrm{A}-\textrm{C})\textrm{cos}(3\psi)+(3\textrm{A}+\textrm{C})\left[\textrm{cos}(\psi)-\sqrt{3}\textrm{sin}(\psi)\right]\\
	2(\textrm{A}-\textrm{C})\textrm{cos}(3\psi)+(3\textrm{A}+\textrm{C})\left[\textrm{cos}(\psi)+\sqrt{3}\textrm{sin}(\psi)\right]\\
	4\textrm{cos}(\psi)\left[-2\textrm{A}+(\textrm{A}-\textrm{C})\textrm{cos}(2\psi)\right]\end{pmatrix}.
	\end{equation}
\end{widetext}

The parameters $\textrm{A}$ and $\textrm{C}$ give the strength of the paths via $\Gamma_5$ and $\Gamma_3$, respectively. They can be assigned to tensor components $\textrm{A}=\chi_\textrm{xxyy}$ and $\textrm{C}=\chi_\textrm{xxxx}$ as in Ref.~\cite{Warkentin18}.

Thus, the THG intensity is calculated according to
\begin{equation}\label{eq.I_THG}
I^{3\omega}(\textrm{A},\textrm{C})\propto|O_\textrm{DDD}(\psi,\textrm{A},\textrm{C})O_\textrm{D}(\varphi)|^2.
\end{equation}
A comparison of the THG rotational anisotropies at the 1S* resonance with simulations using Eq.~\eqref{eq.I_THG} for $\textrm{C}/\textrm{A}=1$ is shown in Fig.~\ref{pic.THG_Scheme+Ani}(b). For the parallel configuration, an isotropic signal intensity is expected from Eq.~\eqref{eq.I_THG} whereas in the crossed configuration no signal is allowed. In the measurement, a slight deviation of the parallel signal from the expected one is observed and furthermore, weak THG intensity with a fourfold pattern is present in the crossed configuration. Deviations from the expected shape might be explained by strain in the sample which can disturb the crystal symmetry, or processes beyond the ED approximation. In Reference~\cite{Warkentin18}, the THG signal in GaAs deviated from the isotropic shape and was fitted with $\textrm{C}/\textrm{A}=0.82$.

\begin{figure}[h]
	\begin{center}
		\includegraphics[width=0.48\textwidth]{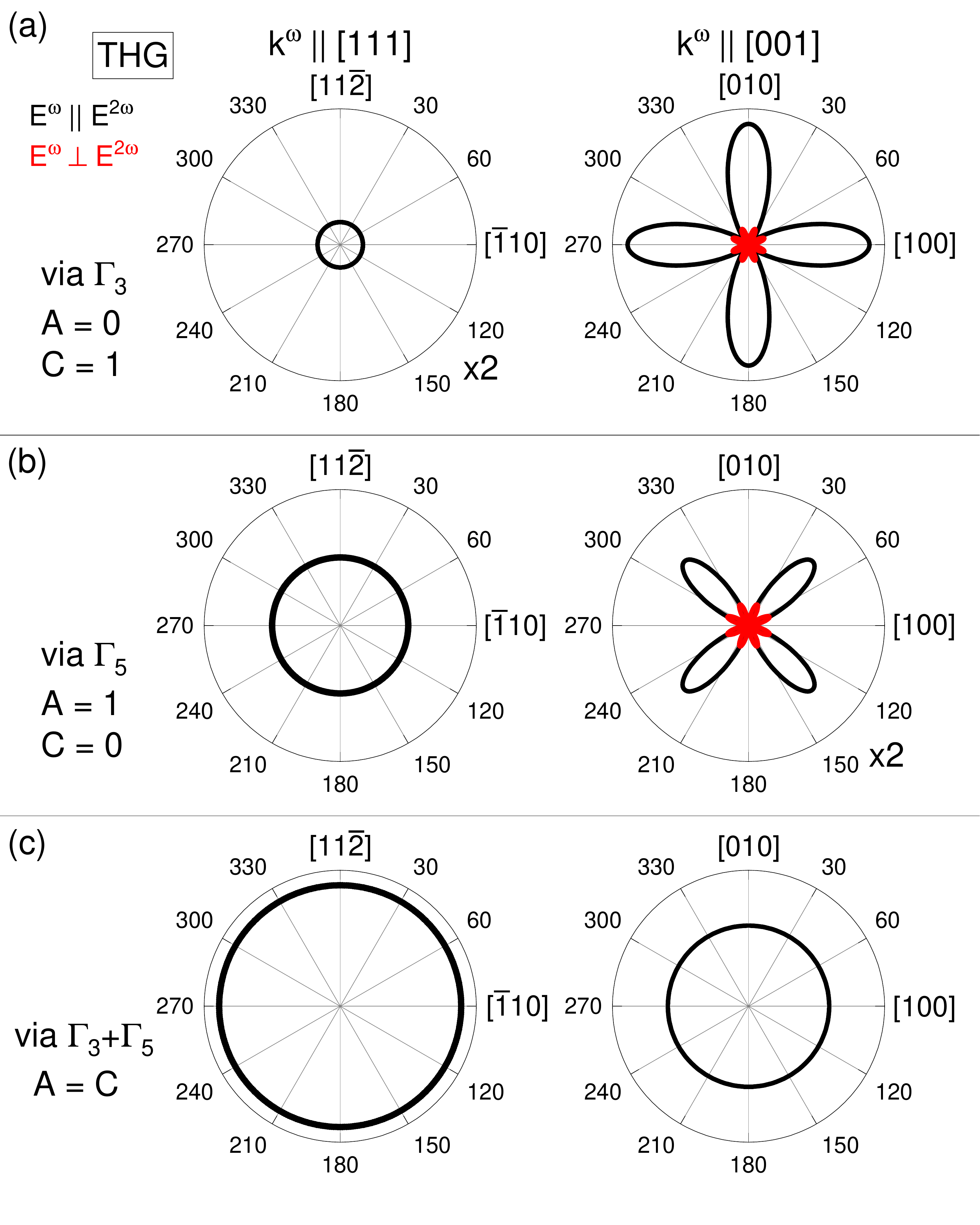}
		\caption{Expected rotational anisotropies for $\textbf{k}^\omega\parallel[111]$ (left side) and $\textbf{k}^\omega\parallel[001]$ (right side) by Eq.~\eqref{eq.I_THG} with $O_\textrm{DDD}$ from Eqs.~\eqref{eq.I_THG} and \eqref{eq.Oddd_001}. Panels show calculation for (a) $\textrm{A}=0$, $\textrm{C}=1$, (b) $\textrm{A}=1$, $\textrm{C}=0$ and (c) $\textrm{A}=\textrm{C}$.}
		\label{pic.THG_path_comparison}
	\end{center}
\end{figure}

In Figures~\ref{pic.THG_path_comparison}(a) and \ref{pic.THG_path_comparison}(b), we show expected anisotropy shapes for the two paths, denoted by A and C, for $\textbf{k}^\omega\parallel[111]$ and $\textbf{k}^\omega\parallel[001]$ crystal direction. Panel (c) shows the interference of both paths with $\textrm{A}=\textrm{C}$. The shapes are calculated by Eq.~\eqref{eq.I_THG} with $O_\textrm{DDD}$ from Eqs.~\eqref{eq.I_THG} and \eqref{eq.Oddd_001}, respectively. Particularly for the [001] direction it is interesting and instructive, that each individual path via either $\Gamma_3$ or $\Gamma_5$ state results in strongly modulated diagrams, whereas their interference for the case $\textrm{A}=\textrm{C}$ results in an isotropic pattern.

\subsection{Fourth harmonic generation}

\begin{figure}[h]
	\begin{center}
		\includegraphics[width=0.48\textwidth]{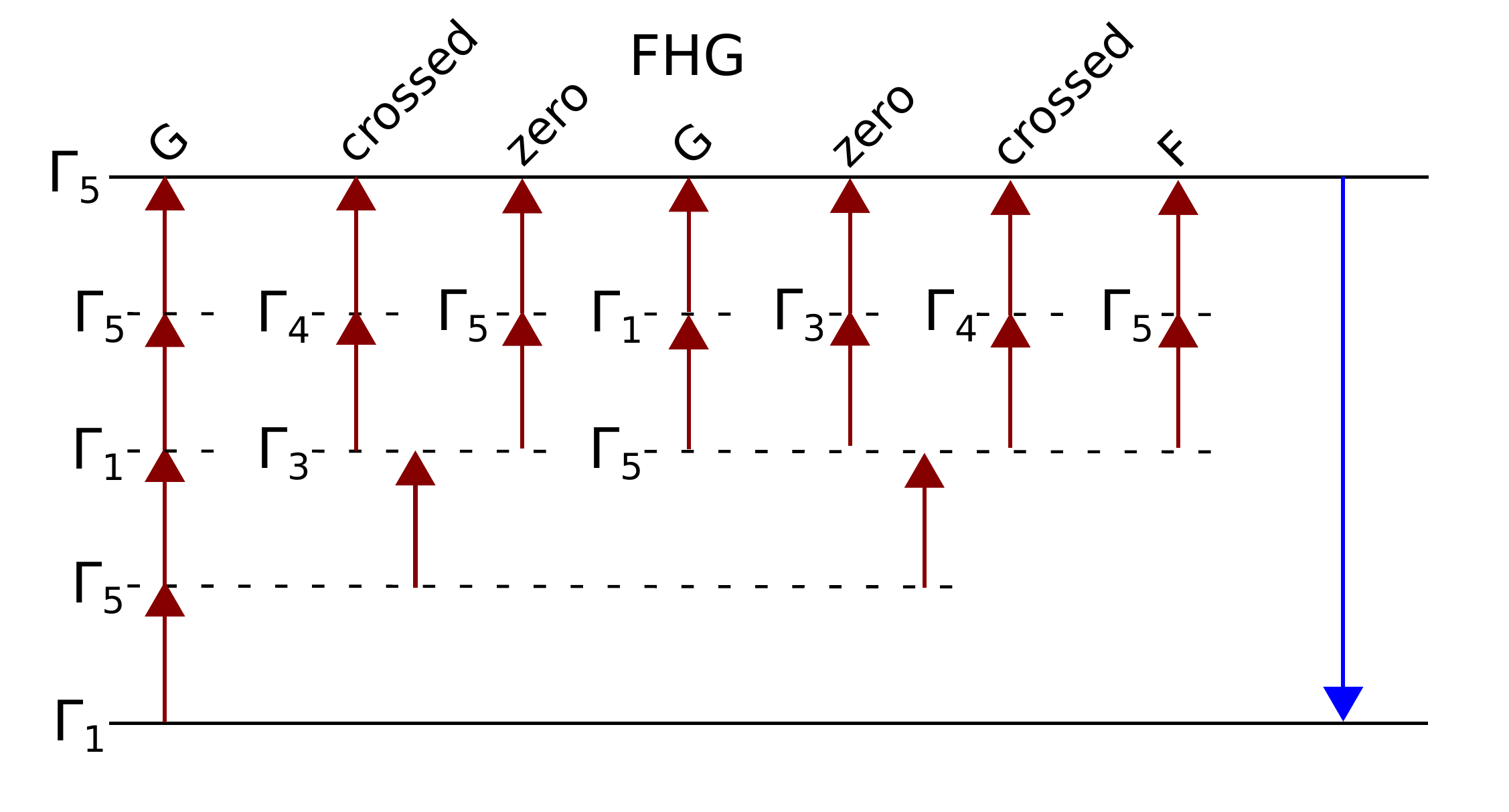}
		\caption{Excitation paths for $\Gamma_5$ excitons in FHG. Dashed lines are a guide to the eye for virtually excited intermediate states with corresponding symmetries. Arrows represent photon transitions of $\Gamma_5$ symmetry. The result of each path is described at the top line: 'F' and 'G' correspond to tensor components of the microscopic analysis. 'zero' means no signal from that path. 'crossed' means only crossed signal from that path.}
		\label{pic.4HG_Scheme}
	\end{center}
\end{figure}

\begin{figure}[h]
	\begin{center}
		\includegraphics[width=0.48\textwidth]{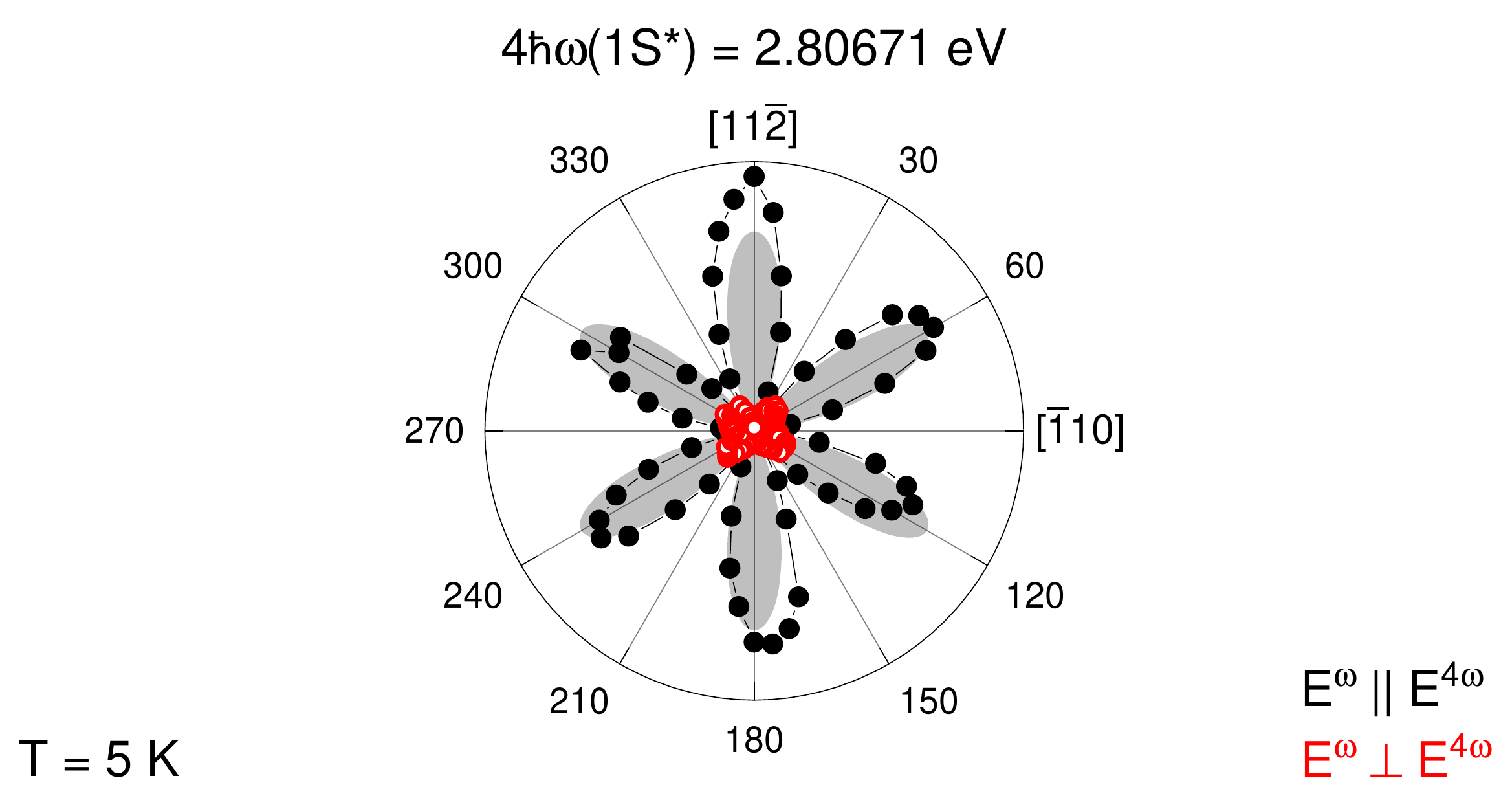}
		\caption{Full black and open red dots give measured FHG data at the 1S* resonance for parallel $\textbf{E}^\omega_4\parallel\textbf{E}^{4\omega}$ and crossed $\textbf{E}^\omega_4\perp\textbf{E}^{4\omega}$ configuration, respectively. The gray shaded area represents the simulation for parallel configuration by Eq.~(\ref{eq.I_4HG}). No signal is expected in crossed configuration.}
		\label{pic.4HG_1S_Ani}
	\end{center}
\end{figure}

In the case of FHG, four photons are involved in the excitation providing a variety of intermediate states and excitation paths. The possible paths for a $\Gamma_5$ exciton in FHG are depicted in Fig.~\ref{pic.4HG_Scheme}. From the symmetry calculation, several paths can be excluded. As shown in Fig.~\ref{pic.4HG_Scheme} the paths that contain $\Gamma_5$ and one $\Gamma_3$ intermediate state (e.g. $\Gamma_5\rightarrow\Gamma_3\rightarrow\Gamma_5\rightarrow\Gamma_5$) give zero FHG signal. Furthermore, those paths which pass a $\Gamma_4$ state result in crossed signal only. Finally, paths that contain either a $\Gamma_1$ (G), or only $\Gamma_5$ (F) intermediate states result in a sixfold signal as obtained in the measurement Fig.~\ref{pic.4HG_1S_Ani}.
The microscopic analysis shows that paths 'G' are much more intense than all others. 'G' can be identified with the tensor component $\chi_\textrm{xxxyz}$ as it requires only one transition through a remote band (as was the case in SHG).

The four photon excitation operator for the light $k$-vector directed along the crystal [111] direction for indistinguishable paths 'G' is given by:
\begin{eqnarray}\nonumber
&&O_\textrm{DDDD}(\textbf{E}^\omega_4, \psi)=\frac{1}{9\sqrt{2}}\\
&&\begin{pmatrix}
\left[1+2\textrm{cos}(2\psi)\right]\left[\sqrt{3}+3\textrm{cot}(\psi)\right]\textrm{sin}^2(\psi)\\
\left[1+2\textrm{cos}(2\psi)\right]\textrm{sin}(\psi)\left[-3\textrm{cos}(\psi)+\sqrt{3}\textrm{sin}(\psi)\right]\\
\sqrt{3}\left[-\textrm{cos}(2\psi)+\textrm{cos}(4\psi)\right]	
\end{pmatrix}.
\end{eqnarray}
Thus, the FHG intensity is calculated from
\begin{equation}\label{eq.I_4HG}
I^{4\omega}\propto|O_\textrm{DDDD}(\psi)O_\textrm{D}(\varphi)|^2.
\end{equation}

A comparison of the FHG rotational anisotropy of the 1S* line with the simulation according to Eq.~(\ref{eq.I_4HG}) is shown in Fig.~\ref{pic.4HG_1S_Ani}. From the simulation the parallel anisotropy is expected to show a sixfold pattern, as was the case in SHG, compare with Fig.~\ref{pic.2PLE+SHG_1S_111_Anis+Simu}(b). In FHG, the crossed configuration is expected to show no signal at all.

\subsection{Rotational anisotropies for various crystal orientations}

\begin{figure*}[t]
	\begin{center}
		\includegraphics[width=0.9\textwidth]{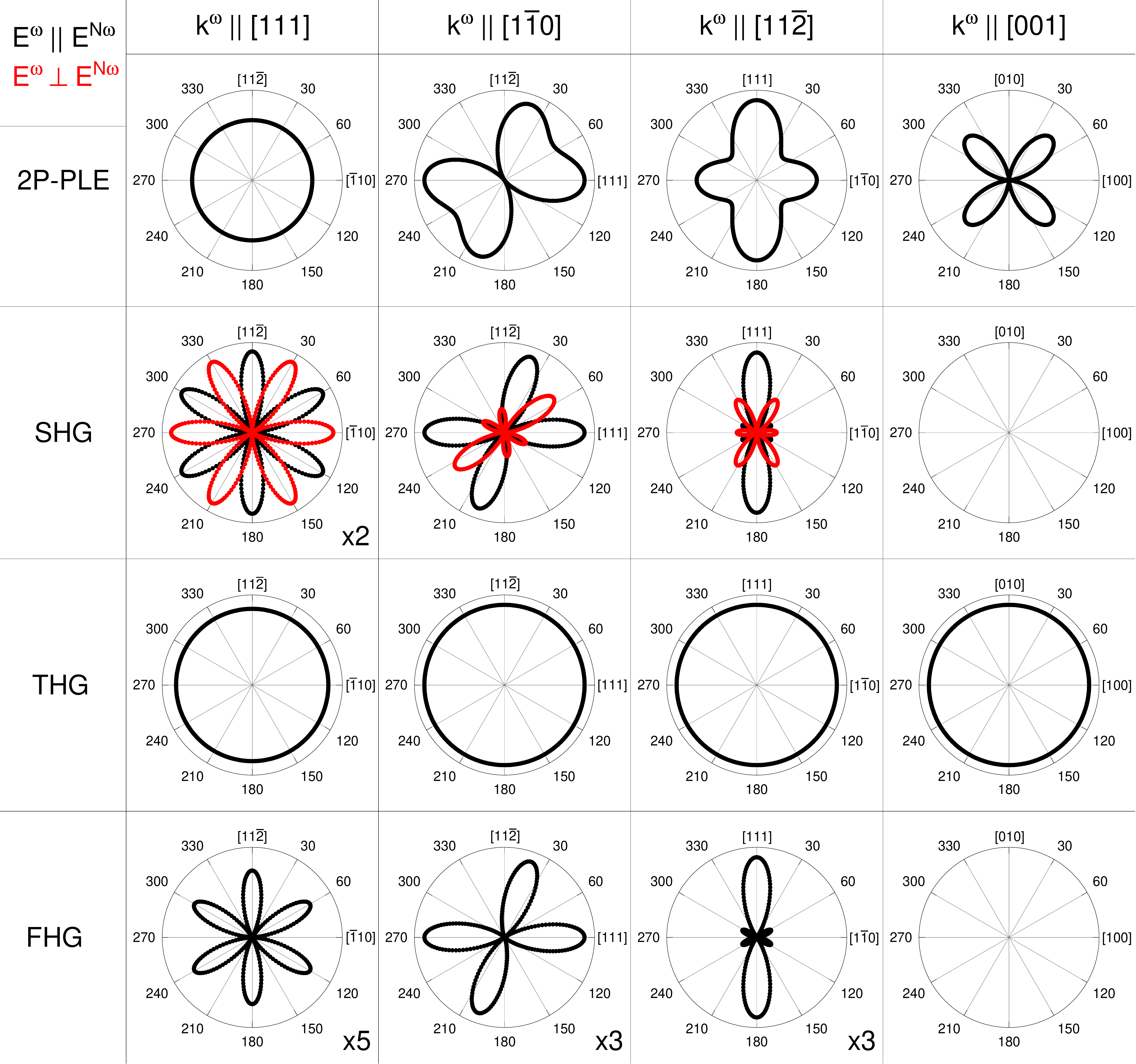}
		\caption{Calculation of rotational anisotropies for S-excitons of $\Gamma_5$ symmetry in 2P-PLE, SHG, THG and FHG for several crystal orientations. Black and red lines give the simulations for parallel $\textbf{E}^\omega\parallel\textbf{E}^{N\omega}$ and crossed $\textbf{E}^\omega\perp\textbf{E}^{N\omega}$ configuration, respectively. THG anisotropies are calculated using $\textrm{A}=\textrm{C}$.}
		\label{pic.ZnSe_Ani_Simulations}
	\end{center}
\end{figure*}

The calculated rotational anisotropies for $S$ excitons of $\Gamma_5$ symmetry in 2P-PLE and optical harmonics generation (up to FHG) for several crystal orientations are given in Fig.~\ref{pic.ZnSe_Ani_Simulations}. THG anisotropies are calculated setting $\textrm{A}=\textrm{C}$. The corresponding excitation operators are presented in the Appendix. One can see that the shape of the rotational anisotropies is very close for SHG and FHG. Two-photon absorption is allowed for all considered orientations. The stricter selection rules for harmonic generation are illustrated by the fact that for a light $k$-vector along the high symmetry direction [001], SHG and FHG are forbidden. Nevertheless, in particular this forbidden orientation allows us to study conveniently the field-induced mechanisms in optical harmonic generation, e.g., magnetic-field-induced signals \cite{Pavlov05, Lafrentz13, Farenbruch20}. For the allowed orientations, the anisotropies present an opportunity to determine the crystal axis by pattern analysis. Furthermore, even a small tilting of the sample from the nominal orientation can be noticed and corrected through anisotropy shapes deviating from the theory. The anisotropies also provide a possibility to resolve state splittings on the order of $\mu$eV through the pattern distortions without the need for high optical resolution \cite{Mund19}. All in all, the rotational anisotropies are in many respects a valuable tool in optical harmonic generation.

\section{Conclusions}
\label{sec_conclusions}

We have extended the recently developed approach of SHG spectroscopy by excitation with spectrally broad femtosecond laser pulses and spectrally-resolved detection for SHG, THG, and FHG studies of exciton-polaritons in the semiconductor ZnSe. The observed spectral shift of the optical harmonic generation signals from the exciton resonance in the reflectivity spectrum of about 3.2~meV is well described by the exciton-polariton dispersion. The experimentally measured rotational anisotropies for light propagation along the [111] crystal axis are in good agreement with the group theory modeling. We provide simulations also for other crystal orientations and show that, e.g., the SHG is symmetry forbidden for [001] orientation, for which therefore application of an external magnetic field is promising to induce magnetic-field-induced SHG signals. The developed approach can be readily extended for other semiconductors and semiconductor heterostructures, e.g., for exciton-polaritons in microcavities in the strong coupling regime.

\section*{ACKNOWLEDGMENTS}
\label{sec_acknowledgments}

The authors are thankful to D. Fr\"ohlich, M. M. Glazov, and E. L. Ivchenko for fruitful discussions. We acknowledge the financial support by the Deutsche Forschungsgemeinschaft through the International Collaborative Research Centre TRR 160 (Project C8) and the Collaborative Research Centre TRR 142 (Project B01).

\section*{APPENDIX: FORMULAS FOR ANISOTROPY SIMULATION}
\label{sec_appendix}

\setcounter{equation}{0}
\renewcommand\theequation{A.\arabic{equation}}

Here, we give the explicit forms of the operators $O_\textrm{D}$ which give the one-photon emission and $O_\textrm{ND}$ for excitation in the different harmonic orders. Formulas are shown for $\textbf{k}^\omega_N\parallel[1\bar{1}0]$, [11$\bar{2}$], and [001]. Note, that $O_\textrm{DD}$ is the same for 2P-PLE and SHG since in both cases excitation is done by two photons.

\subsection{One-photon emission}

One-photon emission operator for $\textbf{k}^{N\omega}\parallel[1\bar{1}0]$:
\begin{equation}
O_\textrm{D}(\textbf{E}^{N\omega}, \varphi)=\frac{1}{\sqrt{6}}\begin{pmatrix}
\textrm{cos}(\varphi)+\sqrt{2}\textrm{sin}(\varphi)\\
\textrm{cos}(\varphi)+\sqrt{2}\textrm{sin}(\varphi)\\
-2\textrm{cos}(\varphi)+\sqrt{2}\textrm{sin}(\varphi)
\end{pmatrix}.
\end{equation}

One-photon emission operator for $\textbf{k}^{N\omega}\parallel[11\bar{2}]$:
\begin{equation}
O_\textrm{D}(\textbf{E}^{N\omega}, \varphi)=\frac{1}{\sqrt{6}}\begin{pmatrix}
\sqrt{2}\textrm{cos}(\varphi)+\sqrt{3}\textrm{sin}(\varphi)\\
\sqrt{2}\textrm{cos}(\varphi)-\sqrt{3}\textrm{sin}(\varphi)\\
\sqrt{2}\textrm{cos}(\varphi)
\end{pmatrix}.
\end{equation}

One-photon emission operator for $\textbf{k}^{N\omega}\parallel[001]$:
\begin{equation}
O_\textrm{D}(\textbf{E}^{N\omega}, \varphi)=\begin{pmatrix}
-\textrm{sin}(\varphi)\\
\textrm{cos}(\varphi)\\
0
\end{pmatrix}.
\end{equation}

\subsection{Two-photon excitation}
	
Two-photon excitation operator for $\textbf{k}^\omega_N\parallel[1\bar{1}0]$:
\begin{eqnarray}\nonumber
&&O_\textrm{DD}(\textbf{E}^\omega_2, \psi)=\frac{1}{6\sqrt{2}}\times\\
&&\begin{pmatrix}
-4\textrm{cos}(2\psi)-\sqrt{2}\textrm{sin}(2\psi)\\
-4\textrm{cos}(2\psi)-\sqrt{2}\textrm{sin}(2\psi)\\
\left[\sqrt{2}\textrm{cos}(\psi)+2\textrm{sin}(\psi)\right]^2
\end{pmatrix}.
\end{eqnarray}

Two-photon excitation operator for $\textbf{k}^\omega_N\parallel[11\bar{2}]$:
\begin{eqnarray}\nonumber
&&O_\textrm{DD}(\textbf{E}^\omega_2, \psi)=\frac{1}{6\sqrt{2}}\times\\
&&\begin{pmatrix}
2\sqrt{2}\textrm{cos}(\psi)\left[\sqrt{2}\textrm{cos}(\psi)-\sqrt{3}\textrm{sin}(\psi)\right]\\
2\sqrt{2}\textrm{cos}(\psi)\left[\sqrt{2}\textrm{cos}(\psi)+\sqrt{3}\textrm{sin}(\psi)\right]\\
-1+5\textrm{cos}(2\psi)
\end{pmatrix}.
\end{eqnarray}

Two-photon excitation operator for $\textbf{k}^\omega_N\parallel[001]$:
\begin{eqnarray}
O_\textrm{DD}(\textbf{E}^\omega_2, \psi)=
\begin{pmatrix}
0\\
0\\
-\sqrt{2}\textrm{cos}(\psi)\textrm{sin}(\psi)
\end{pmatrix}.
\end{eqnarray}

\begin{widetext}
\subsection{Three-photon excitation}

Three-photon excitation operator for $\textbf{k}^\omega_N\parallel[1\bar{1}0]$:
\begin{eqnarray}\nonumber
&&O_\textrm{DDD}(\textbf{E}^\omega_3, \psi, \textrm{A}, \textrm{C})=\frac{1}{12\sqrt{3}}\times\\
&&\begin{pmatrix}
\left[\sqrt{2}(5\textrm{A}-\textrm{C})\textrm{cos}^3(\psi)-2(3\textrm{A}+\textrm{C})\textrm{cos}^2(\psi)\textrm{sin}(\psi)+4\sqrt{2}\textrm{C}\textrm{cos}(\psi)\textrm{sin}^2(\psi)-8\textrm{A}\textrm{sin}^3(\psi)\right]\\
\left[\sqrt{2}(5\textrm{A}-\textrm{C})\textrm{cos}^3(\psi)-2(3\textrm{A}+\textrm{C})\textrm{cos}^2(\psi)\textrm{sin}(\psi)+4\sqrt{2}\textrm{C}\textrm{cos}(\psi)\textrm{sin}^2(\psi)-8\textrm{A}\textrm{sin}^3(\psi)\right]\\
2\left[\sqrt{2}\textrm{cos}(\psi)+\textrm{sin}(\psi)\right]\left[-3\textrm{A}-\textrm{C}+(\textrm{A}-\textrm{C})(\textrm{cos}(2\psi)+2\sqrt{2}\textrm{sin}(2\psi))\right]
\end{pmatrix}.
\end{eqnarray}

Three-photon excitation operator for $\textbf{k}^\omega_N\parallel[11\bar{2}]$:
\begin{eqnarray}\nonumber
&&O_\textrm{DDD}(\textbf{E}^\omega_3, \psi, \textrm{A}, \textrm{C})=\frac{1}{24\sqrt{3}}\times\\
&&\begin{pmatrix}
2\left[2\textrm{cos}(\psi)+\sqrt{6}\textrm{sin}(\psi)\right]\left[4\textrm{A}\textrm{cos}^2(\psi)+(3\textrm{A}+\textrm{C})\textrm{sin}^2(\psi)+\sqrt{6}(-\textrm{A}+\textrm{C})\textrm{sin}(2\psi)\right]\\
\left[2\textrm{cos}(\psi)-\sqrt{6}\textrm{sin}(\psi)\right]\left[7\textrm{A}+\textrm{C}+(\textrm{A}-\textrm{C})(\textrm{cos}(2\psi)+2\sqrt{6}\textrm{sin}(2\psi))\right]\\
-4\textrm{cos}(\psi)\left[-5\textrm{A}+\textrm{C}+(\textrm{A}-\textrm{C})\textrm{cos}(2\psi)\right]
\end{pmatrix}.
\end{eqnarray}

Three-photon excitation operator for $\textbf{k}^\omega_N\parallel[001]$:
\begin{eqnarray}\label{eq.Oddd_001}
O_\textrm{DDD}(\textbf{E}^\omega_3, \psi, \textrm{A}, \textrm{C})=\frac{1}{6}
\begin{pmatrix}
\left[3\textrm{A}+\textrm{C}+3(\textrm{A}-\textrm{C})\textrm{cos}(2\psi)\right]\textrm{sin}(\psi)\\
\textrm{cos}(\psi)\left[3\textrm{A}+\textrm{C}+3(-\textrm{A}+\textrm{C})\textrm{cos}(2\psi)\right]\\
0
\end{pmatrix}.
\end{eqnarray}


\subsection{Four-photon excitation}

Four-photon excitation operator for $\textbf{k}^\omega_N\parallel[1\bar{1}0]$:
\begin{equation}
O_\textrm{DDDD}(\textbf{E}^\omega_4, \psi)=\frac{1}{24\sqrt{6}}
\begin{pmatrix}
-2\left[-2\textrm{cos}(\psi)+\sqrt{2}\textrm{sin}(\psi)\right]^3\left[\textrm{cos}(\psi)+\sqrt{2}\textrm{sin}(\psi)\right]\\
-2\left[-2\textrm{cos}(\psi)+\sqrt{2}\textrm{sin}(\psi)\right]^3\left[\textrm{cos}(\psi)+\sqrt{2}\textrm{sin}(\psi)\right]\\
9+7\textrm{cos}(4\psi)+4\sqrt{2}\textrm{sin}(4\psi)
\end{pmatrix}.
\end{equation}

Four-photon excitation operator for $\textbf{k}^\omega_N\parallel[11\bar{2}]$:
\begin{equation}
O_\textrm{DDDD}(\textbf{E}^\omega_4, \psi)=-\frac{1}{36}
\begin{pmatrix}
\left[1+5\textrm{cos}(2\psi)\right]\textrm{sin}(\psi)\left[-3\textrm{cos}(\psi)+\sqrt{6}\textrm{sin}(\psi)\right]\\
\left[1+5\textrm{cos}(2\psi)\right]\textrm{sin}(\psi)\left[3\textrm{cos}(\psi)+\sqrt{6}\textrm{sin}(\psi)\right]\\
\sqrt{6}\textrm{sin}^2(\psi)\left[1+5\textrm{cos}(2\psi)\right]
\end{pmatrix}.
\end{equation}
\end{widetext}

Four-photon excitation operator for $\textbf{k}^\omega_N\parallel[001]$:
\begin{equation}
O_\textrm{DDDD}(\textbf{E}^\omega_4, \psi)=
\begin{pmatrix}
0\\
0\\
0
\end{pmatrix}.
\end{equation}



\end{document}